\documentclass[twocolumn]{aastex631}

\usepackage{natbib}
\usepackage[style=american]{csquotes}
\usepackage[bulgarian,english]{babel}
\usepackage[T2A]{fontenc}

\usepackage{scrextend}
\let\orgautoref\autoref
\renewcommand{\autoref}
        {\def\equationautorefname{Eq.}%
         \def\figureautorefname{Fig.}%
         \def\sectionautorefname{Sect.}%
         \def\subsectionautorefname{Sect.}%
         \def\subsubsectionautorefname{Sect.}%
         \orgautoref}

\shorttitle{Revisiting the HD\,33142 system}
\shortauthors{Trifonov et al.}
\graphicspath{{./}{figures/}}

\usepackage{changepage} 
\usepackage{rotating}

\begin{document}


\title{A new third planet and the dynamical architecture of the HD\,33142 planetary system\footnote{Based on observations collected at the European Organization for Astronomical Research in the Southern Hemisphere under ESO programmes 60.A-9700, 60.A-9036, 097.C-0090, 0100.C-0414, 0101.C-0232, 0102.C-0338, and MPG programmes 088.C-0892,  099.A-9009, 0100.A-9006.}}

\correspondingauthor{Trifon Trifonov}
\email{trifonov@mpia.de}

\author[0000-0002-0236-775X]{Trifon Trifonov}
\affiliation{Max-Planck-Institut für Astronomie,
              Königstuhl 17,
              69117 Heidelberg, Germany}
\affiliation{Department
 of Astronomy, Sofia University ``St Kliment Ohridski'', 5 James Bourchier Blvd, BG-1164 Sofia, Bulgaria}

\author{Anna Wollbold}
\affiliation{Max-Planck-Institut für Astronomie,
              Königstuhl 17,
              69117 Heidelberg, Germany}

\author[0000-0002-1765-9907]{Martin Kürster}
\affiliation{Max-Planck-Institut für Astronomie,
              Königstuhl 17,
              69117 Heidelberg, Germany}

\author[0000-0003-3130-2768]{Jan Eberhardt}
\affiliation{Max-Planck-Institut für Astronomie,
              Königstuhl 17,
              69117 Heidelberg, Germany}

\author[0000-0002-1166-9338]{Stephan Stock}
\affiliation{Landessternwarte, Zentrum für Astronomie der Universtät Heidelberg,
              Königstuhl 12,
              69117 Heidelberg, Germany}

\author[0000-0002-1493-300X]{Thomas Henning}
\affiliation{Max-Planck-Institut für Astronomie,
              Königstuhl 17,
              69117 Heidelberg, Germany}

\author[0000-0002-0460-8289]{Sabine Reffert}
\affiliation{Landessternwarte, Zentrum für Astronomie der Universtät Heidelberg,
              Königstuhl 12,
              69117 Heidelberg, Germany}
     
\author[0000-0003-1305-3761]{R. Paul Butler}
\affiliation{Earth and Planets Laboratory, Carnegie Institution for Science, Washington DC 20015, USA}

\author{Steven S. Vogt}
\affiliation{UCO/Lick Observatory, Department of Astronomy and Astrophysics, University of California at Santa Cruz, Santa Cruz, CA 95064, USA}

\author[0000-0003-1242-5922]{Ansgar Reiners}
\affiliation{Institut f\"ur Astrophysik und Geophysik, Georg-August-Universität,
              Friedrich-Hund-Platz 1,
              37077 Göttingen, Germany}

\author[0000-0003-1930-5683]{Man Hoi Lee}
\affiliation{Department of Earth Sciences, The University of Hong Kong,
Pokfulam Road,
Hong Kong}
\affiliation{Department of Physics, The University of Hong Kong,
Pokfulam Road,
Hong Kong }

\author[0000-0002-8868-7649]{Bertram Bitsch}
\affiliation{Max-Planck-Institut für Astronomie, Königstuhl 17, 69117 Heidelberg, Germany}

\author[0000-0002-6532-4378]{Mathias Zechmeister}
\affiliation{Institut für Astrophysik, Georg-August-Universität,
              Friedrich-Hund-Platz 1,
              37077 Göttingen, Germany}

\author[0000-0003-0650-5723]{Florian Rodler}
\affiliation{European Southern Observatory (ESO), Alonso de Cordova 3107, 
              Vitacura, Santiago de Chile, Chile}
              
\author[0000-0002-6859-0882]{Volker Perdelwitz}
\affiliation{Department of Physics, Ariel University, Ariel 40700, Israel}

\author[0000-0003-3757-1440]{Lev Tal-Or}
\affiliation{Department of Physics, Ariel University, Ariel 40700, Israel}
\affiliation{Astrophysics Geophysics And Space Science Research Center, Ariel University, Ariel 40700, Israel}

\author[0000-0002-0993-6089]{Jan Rybizki}
\affiliation{Max-Planck-Institut für Astronomie, Königstuhl 17, 69117 Heidelberg, Germany}

\author[0000-0002-3662-9930]{Paul Heeren}
\affiliation{Landessternwarte, Zentrum für Astronomie der Universtät Heidelberg,
              Königstuhl 12,
              69117 Heidelberg, Germany}

\author[0000-0001-8627-9628]{Davide Gandolfi}
\affiliation{Dipartimento di Fisica, Università di Torino,
via P. Giuria 1, 10125 Torino, Italy}

\author[0000-0003-0563-0493]{Oscar Barragán}
\affiliation{Sub-department of Astrophysics, Department of Physics, University of Oxford, Oxford OX1 3RH, UK }

\author[0000-0002-4199-6356]{Olga Zakhozhay}
\affiliation{Max-Planck-Institut für Astronomie, Königstuhl 17, 69117 Heidelberg, Germany}
\affiliation{Main Astronomical Observatory, National Academy of Sciences of Ukraine, Kyiv 03143, Ukraine}

\author[0000-0001-8128-3126]{Paula Sarkis}
\affiliation{Max-Planck-Institut für Astronomie, Königstuhl 17, 69117 Heidelberg, Germany}

\author[0000-0003-1566-7740]{Marcelo Tala Pinto}
\affiliation{Landessternwarte, Zentrum für Astronomie der Universtät Heidelberg, Königstuhl 12,
              69117 Heidelberg, Germany}

\author[0000-0002-0436-7833]{Diana Kossakowski}
\affiliation{Max-Planck-Institut für Astronomie, Königstuhl 17, 69117 Heidelberg, Germany}

\author{Vera Wolthoff}
\affiliation{Landessternwarte, Zentrum für Astronomie der Universtät Heidelberg,
              Königstuhl 12,
              69117 Heidelberg, Germany}

\author[0000-0002-1440-3666]{Stefan S. Brems}
\affiliation{Landessternwarte, Zentrum für Astronomie der Universtät Heidelberg,
              Königstuhl 12,
              69117 Heidelberg, Germany}

\author[0000-0002-8569-7243]{Vera Maria Passegger}
\affiliation{Hamburger Sternwarte, Gojenbergsweg 112, 21029 Hamburg, Germany
}
\affiliation{Homer L. Dodge Department of Physics and Astronomy, University of Oklahoma, 440 West Brooks Street, Norman, OK 73019,
United States of America
}




\begin{abstract}

Based on recently-taken and archival HARPS, FEROS and HIRES radial velocities (RVs), we present evidence for a new planet orbiting the first ascent red giant star HD\,33142 (with an improved mass estimate of 1.52$\pm$0.03\,M$_\odot$), already known to host two planets. 
We confirm the Jovian mass planets HD\,33142\,b and c with periods of
$P_{\rm b}$ = 330.0$_{-0.4}^{+0.4}$\,d and $P_{\rm c}$ = 810.2$_{-4.2}^{+3.8}$\,d and minimum dynamical masses of $m_{\rm b}\sin{i}$ = 1.26$_{-0.05}^{+0.05}$\,M$_{\rm Jup}$ and $m_{\rm c}\sin{i}$ = 0.89$_{-0.05}^{+0.06}$\,M$_{\rm Jup}$. Furthermore, our periodogram analysis of the precise RVs shows strong evidence for a short-period Doppler signal in the residuals of a two-planet Keplerian fit,
which we interpret as a third, Saturn-mass planet with $m_\mathrm{d}\sin{i}$ = 0.20$_{-0.03}^{+0.02}$\, M$_{\rm Jup}$ on a close-in orbit with an orbital period of $P_{\rm d}$ =89.9$_{-0.1}^{+0.1}$\,d. 
We study the dynamical behavior of the three-planet system configurations with an N-body integration scheme,
finding it long-term stable with the planets alternating between low and moderate eccentricities episodes.
We also performed N-body simulations, including stellar evolution and second-order dynamical effects such as planet-stellar tides and stellar mass-loss on the way to the white dwarf phase. We find that planets HD\,33142 b, c \& d are likely to be engulfed near the tip of the red giant branch phase due to tidal migration.
These results make the HD\,33142 system an essential benchmark for the planet population statistics of the multiple-planet systems found around evolved stars.\looseness=-4
\end{abstract}

\keywords{planetary system ---  stars: late-type --- planets and satellites: dynamical evolution and stability}


\section{Introduction} \label{sec:intro}

Since the first detection of an exoplanet around a solar-type star in 1995 \citep{Mayor1995}, the number of confirmed exoplanets has increased rapidly to over 5000\footnote{NASA Exoplanet Archive: \url{https://exoplanetarchive.ipac.caltech.edu/} (31.05.2022)}, from which the majority have been discovered using the transit method, followed by the radial velocity (RV) method. The RV technique is the best way to determine the orbital architecture of multiple-planet systems, which is important to improve our understanding of planet formation and orbital evolution. However, a sparse phase coverage of the orbit together with data errors can lead to model ambiguities, as for example when two planets near a 2:1 period ratio are mistaken for a single planet on an eccentric orbit \citep{Anglada2010,Wittenmyer2013,Kuerster2015,Boisvert2018,Hara2019}. \cite{Kuerster2015} used the Exoplanet Orbit Database (EOD; \url{http://www.exoplanets.org/}) to determine that for 74\% of 254 putative single-planet systems, detected via the radial velocity technique, two planets in a 2:1 resonance are also a viable model. This model ambiguity was the motivation to start a survey with HARPS and FEROS for follow-up observations of stars previously reported to host a single eccentric planet, but so far with sparse RV sampling. The goal of this program is to look for additional planets in these systems.

This paper presents a new orbital analysis of the HD\,33142 system, which was part of our HARPS and FEROS multiple-planet system search. HD\,33142 was included in our survey, known to host a moderately eccentric ($e_b$ $\approx$ 0.22) Jovian-mass planet HD\,33142\,b, orbiting at a period of $\sim$ 330\,d \citep{Johnson2011}. While we were gathering observational data of HD\,33142, additional HIRES RVs made it possible for \citet{Bryan2016}, and then \citet{Luhn2019} to announce the presence of a second long-period planet HD\,33142\,c, with a period of $\sim 800$\,d.  
Based on our new data, we add the discovery of HD\,33142\,d, a Saturn-mass planet with $P_\mathrm{d} \sim 89.9\,\mathrm{d}$, and refined the orbital parameters for HD\,33142\,b and HD\,33142\,c. We also provide an analysis of the orbital dynamics and stability of the three-planet system. 
Furthermore, we present N-body simulation results, including planet-stellar tides, stellar evolution and  mass-loss from the red giant branch (RGB) to the white dwarf (WD) phase.

This paper is organized as follows: 
In \autoref{Sec2}, we introduce the planet host HD\,33142 and the two known planets. In  \autoref{Sec3} we present the new radial velocity and spectral activity data obtained by HARPS and FEROS, as well as the archival HIRES data. \autoref{Sec4} summarizes our data analysis and our results, and \autoref{Sec5} presents the pertinent discussion. Finally, \autoref{Sec6} contains our conclusions.

\section{Overview of the HD\,33142 system}
\label{Sec2}

\subsection{Stellar parameters}
\label{Sec2.1}

\begin{figure}
    \centering
    \includegraphics[width=8.5cm]{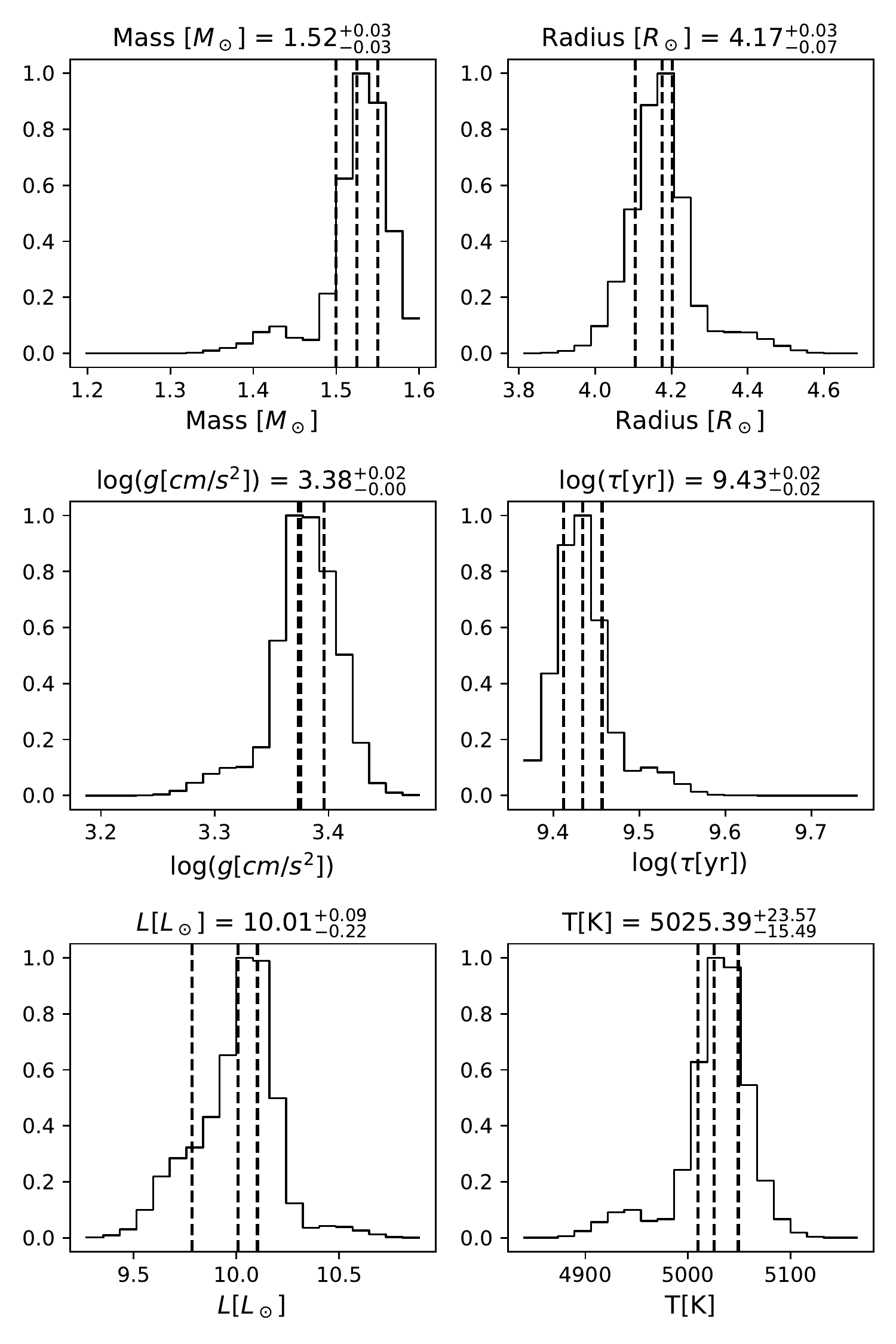}
    \caption{Probability density functions of stellar parameters of HD\,33142 derived by applying the Bayesian inference scheme by \cite{Stock2018} and using the parallax, $B$ and $V$ photometry, and the spectroscopic metallicity estimate provided in Table~\ref{table_st_param}. The vertical dashed lines show the 0.16, 0.5 (median) and 0.84 quantiles of the posterior samples.\looseness=-4}
\label{fig:Parameters} 
\end{figure}

HD\,33142 (HIP\,23844, BD-14 1051) has a $V$-band magnitude of $7.96$\,mag and a color index of $B-V = 0.935$\,mag. The {\it Gaia} DR2 parallax is 
$\pi = 8.2103 \pm$ 0.0394\,mas, corresponding to a distance of $d$ = 121.80 $\pm$ 0.58\,pc \citep{Gaia_Collaboration2018b}. \citet{Houk1988} 
measured the spectral type to be K0\,III, which makes HD\,33142 a giant star, but it is also referred to as a subgiant in various works \citep[e.g.,][]{Ghezzi2018,Luhn2019}. Therefore, we determine the stellar mass $M_\ast $, radius $R_\ast $, luminosity $L_\ast$, effective temperature $T_\mathrm{eff}$, surface gravity $\log\,g_\ast$ and age $\tau_\ast$, as well as the most probable current evolutionary stage, following the Bayesian inference scheme used in \cite{Stock2018}. The method of \cite{Stock2018} uses priors and the log-likelihood of observable parameters given a grid of stellar evolution models to compute weighted probability density functions (PDFs) of stellar parameters that are not directly observable, such as stellar mass, from which these parameters can then be statistically inferred. We show the PDFs of the derived stellar parameters of HD\,33142 in Fig.~\ref{fig:Parameters}. The code uses the stellar models based on the PAdova and TRieste Stellar Evolution  Code \citep[PARSEC; ][]{Bressan2012}. The method by \cite{Stock2018} has several advantages compared to other mass estimation methods that are based on photometric and spectroscopic observables. First, it incorporates prior information in the form of the stellar evolutionary time and an initial mass function. Second, Bayesian inference is performed in the plane of the astrometric HR diagram \citep{Arenou1999}, which has a color as the abscissa and the astrometry-based luminosity (ABL) in a given photometric band as the ordinate. The ABL is a quantity in which the trigonometric parallax is linearly incorporated, allowing an unbiased comparison of stellar positions with evolutionary models in cases where the parallax error dominates. Third, the code is capable of deriving the most probable current evolutionary stage of the studied star. 
\cite{Stock2018} showed that the method reliably reproduces model-independent asteroseismic mass measurements and evolutionary stages of sub-giant and giant stars (both red-giant branch and horizontal branch/red clump stars). This was again confirmed by \cite{Malla2020} who noted that the method by \cite{Stock2018}\footnote{A \texttt{Python} implementation of the original, unpublished \texttt{IDL} code used by \cite{Stock2018} is available in \url{https://github.com/StephanStock/SPOG/}}, showed the smallest mass offset among their comparisons of spectroscopic mass estimates to asteroseismic mass estimates. 
Regarding HD\,33142, we find it most likely to be a giant star at the beginning of the red giant branch phase. Table \ref{table_st_param} summarizes the main parameters of the star. Its mass and radius are $M_{*}$ = 1.52$^{+0.03}_{-0.03}$\,$M_{\odot}$ and $R_{*}$ = 4.17$^{+0.03}_{-0.07}$\,$R_{\odot}$, respectively. Note that we have neglected any extinction estimate for HD\,33142 as the star is relatively nearby, and extinction is often a very uncertain parameter for an individual star. As an additional check of the reliability of our parameters, we compared the derived radius, luminosity and effective temperature by the Bayesian inference method with {\it Gaia} DR2 estimates \citep[$R_{Gaia}=4.14^{+0.08}_{-0.07}$\,R$_\odot$, $L_{Gaia}=9.94^{+0.07}_{-0.07}$\,L$_\odot$, $T_{Gaia}=5040^{+42}_{-53}$\,K; ][]{Gaia_Collaboration2018b} and found that they are consistent within $1\,\sigma$.

\subsection{Stellar rotation}
\label{Sec2.2}

The rotational velocity of HD\,33142 was first determined by \citet{Johnson2011} to be $v\sin{i} = 2.9\pm 0.5\,\mathrm{km\,s}^{-1}$ 
by analyzing a spectrum from the HIRES instrument \citep{Vogt1994} 
with the Spectroscopy Made Easy (SME) package \citep{Valenti1996}.  \citet{Jofre2015} determined this value to be $v\sin{i} = 1.61\pm 0.22\, {\mathrm{km\,s}}^{-1}$ using spectra from the FEROS instrument \citep{Kaufer1999}. They developed a task in IRAF \citep{Tody1986} to automatically compute the $v\sin{i}$ from the FWHM of 13 isolated iron lines, and then averaged the values. \citet{Ghezzi2018} calculated a value of $v\sin{i} = 1.2\pm 0.8\, {\mathrm{km\,s}}^{-1}$ based on spectral synthesis with MOOG\footnote{ \url{http://www.as.utexas.edu/\string~chris/moog.html}} \citep{s1973} of one Fe I line using HIRES spectra. These values are compiled in Table \ref{table_st_param}.

\begin{table}
\caption{Stellar parameters of HD\,33142.} %

\begin{adjustwidth}{-4.8cm}{}
\resizebox{0.72\textheight}{!}{\begin{minipage}{\textwidth}

\label{table_st_param}
\centering
\begin{tabular}{l r r r r}
\hline
\hline

\noalign{\smallskip}
                Parameter           & HD\,33142        & Reference                  \\
 \hline
\noalign{\smallskip} 
Spectral type                      &  K0\,III      &     \citet{Houk1988}              \\
$V$ [mag]                     & 7.96$\pm$0.01     &      \citet{ESA1997} \\
$B-V$ [mag]                     & 0.953$\pm$0.013  & \citet{ESA1997} \\
$\pi $ [mas]                    & 8.272$\pm0.020$      &    \cite{Gaia2021}    \\

[Fe/H]    &0.06 $\pm0.01$  & \\ 

\hline
\noalign{\smallskip} 
                
$L$ [L$_\odot$] & 10.01$^{+0.09}_{-0.22}$ &  This work \\
$T_{\rm eff}$ [K]             & 5025.4$^{+23.6}_{-15.5}$     &  This work                 \\
$\log{g}$ $[\mathrm{cm\,s}^{-2}]$                    & 3.375$^{+0.021}_{-0.002}$  & This work \\
$R_\star$ [R$_\odot$]         & 4.17$^{+0.03}_{-0.07}$ &  This work         \\
$M_\star$ [M$_\odot$]         & 1.52$^{+0.03}_{-0.03}$  &  This work        \\
$\tau $ [Gyr] & $2.72^{+0.14}_{-0.13}$ & This work \\

\hline
\noalign{\smallskip} 

$v\sin{i}$    [${\mathrm{km\,s}}^{-1}$]            & $1.2\pm 0.8$ & \citet{Ghezzi2018} \\
            & $1.61\pm 0.22$ &  \citet{Jofre2015} \\
            & $2.9\pm 0.5$ & \citet{Johnson2011} \\
             &
$\le 2$ & This work  \\
$P_\mathrm{rot}/\sin{i}$   [d]$^{a}$              & $106$ & This work \\
\noalign{\smallskip}
\hline
\hline
\end{tabular}



 \end{minipage}}
 \end{adjustwidth}
\tablecomments{ a -- $P_\mathrm{rot}/\sin{i}$ value is derived from the $v\sin{i}$'s using Eq.~(1). For an upper and a lower limit, see text for details.}
\end{table}

We also determined the $v\sin{i}$ of HD\,33142 from the combined HARPS spectra, i.e., from the instrument with the highest resolution among those from which we have data.  In effect, we compared the width of absorption lines with that of lines from the slowly rotating K2III star HD\,178484 observed with the same instrument.  This way we overcame systematics from ill-determined broadening effects such as convection on the $1\,\mathrm{km}^{-1}$ scale \citep[see~e.g.,][]{2012AJ....143...93R}. 
As no rotational broadening is detected, we provide an upper
limit of $v\sin{i}\le 2\,\mathrm{km\,s}^{-1}$ in accordance with the relatively high intrinsic broadening of the lines of giant stars.  This value is also included in \autoref{table_st_param} and is the adopted limit of  $v\sin{i}$ in our work.

The $v\sin{i}$ values from the literature are comparable to the upper limit we found and have quite small uncertainties. We interpret the three literature values as formal fit optima depending on the applied method and not considering systematics. For example, the value from \citet{Jofre2015}, which is the one with the smallest formal uncertainty and therefore offering itself as the value of choice, is based on FEROS spectra with less than half the resolution of our HARPS spectra ($R=48,000$ vs.~$R=115,000$), and from Fig.\,9 in that paper, we can estimate that the typical $v\sin{i}$ uncertainty is considerably larger than the $0.22\,\mathrm{km\,s}^{-1}$ these authors provide. Similar arguments apply for the value from \citet{Ghezzi2018}, whereas the \cite{Johnson2011} value is larger than our upper limit by almost twice its quoted error and therefore unlikely to be true.

Using our limit on $v\sin{i}$, we can derive an estimate of the stellar rotation period $P_{\mathrm{rot}}$ from 
\begin{equation}
\frac{P_{\mathrm{rot}}}{\sin{i}}= \frac{2 \pi R_\star }{v \sin{i}}
\end{equation}
with $R_\star = 4.17\,R_\odot $ the stellar radius (see Table \ref{table_st_param}). From Eq.~(1) we find a value of $P_\mathrm{rot}/\sin{i}=106\,\mathrm{d}$.
Unfortunately, this estimate for $P_{\mathrm{rot}}$ from a limit in 
$v\sin{i}$ is neither a lower nor an upper limit, which can be seen as follows.  On the one hand, since the unknown inclination $i$ is in the range $0^\circ \le i \le 90^\circ $, the true period can be shorter than the estimate $P_{\mathrm{rot}}/\sin{i}$, hence $P_\mathrm{rot} \le 106\,\mathrm{d}$.  On the other hand,
our $v\sin{i}$ value is best considered as an upper limit (of the true $v\sin{i}$) and the estimate of the stellar rotation period $P_{\mathrm{rot}}/\sin{i}$ is therefore a lower limit to the period, hence $P_{\mathrm{rot}} \ge 106\,\mathrm{d}$.

 We also inspected the archival {\texttt Hipparcos} photometry \citep{VanLeeuwen2007} and we find no sign of periodicity at longer timescales, which could be associated with the stellar rotation. In summary, our attempts to estimate the true rotation period remain inconclusive, and it is not possible to decide whether any of the found periodic signals coincides with the rotation period.

\subsection{Planetary system}
\label{Sec2.3}

\citet{Johnson2011} announced HD\,33142\,b, a Jovian mass planet with an orbital period of $\sim$ 330\,d. \citet{Bryan2016}  announced HD\,33142\,c with an orbital period of $\sim$ 830\,d and a mass of $m_\mathrm{c}\sin{i} \sim 6\,M_{\mathrm{Jup}}$\footnote{We believe that this literature minimum mass estimate of HD\,33142\,c in  \citet{Bryan2016} is a result of a typo in their Table 5, or a typo in their adopted stellar mass of HD\,33142.}. \citet{Luhn2019} also announced a possible second Jovian planet, HD\,33142\,c. This last work, provides the following values for the minimum masses and orbital periods of the two planets b and c, $m_b \sin i = 1.385\pm 0.064\,\mathrm{M}_\mathrm{Jup}$ and $P_b = 326.0\pm 1.2\,\mathrm{d}$ and $m_c \sin i = 0.62\pm 0.11\,\mathrm{M_\mathrm{Jup}}$ and $P_c = 809\pm 26\,\mathrm{d}$. 

HD\,33142 was observed by the Transiting Exoplanet Survey Satellite \citep[$TESS$,][]{Ricker2015} in sectors 5 and 32. We inspected the $TESS$ light curves and found no transit signals. Overall, the $TESS$ light-curves are rather constant, suggesting a photometrically quiet star.

In this paper (Section \ref{Sec4}), we present the independent discovery of HD\,33142\,c in our data with an orbital period of $\sim 810$\,d and a mass of $m_\mathrm{c}\sin{i} \sim 0.88\,M_{\mathrm{Jup}}$.
We additionally find strong evidence for the presence of HD\,33142\,d; a Saturn-mass, short-period companion which is evident in the combined HIRES, FEROS, and HARPS data sets.

\section{Data}
\label{Sec3}

In this work, we combine new observations of HD\,33142 taken with the HARPS \citep[][]{Mayor2003} and FEROS \citep[][]{Kaufer1999} spectrographs with archival and literature data from the same instruments and from the Keck HIRES spectrograph \citep{Vogt1994}. 
For our analysis, a total of 98 nightly-binned RVs were used, which cover a temporal baseline of over 12 years.

\subsection{FEROS Data}
\label{Sec3.1}  

 The Fiber-fed Extended Range Optical Spectrograph \citep[FEROS;][]{Kaufer1999} is an \'echelle spectrograph installed at the 2.2 meter MPG/ESO\footnote{MPG = {\em Max-Planck-Gesellschaft} = Max Planck Society; ESO = European Southern Observatory).} telescope at ESO's La Silla Observatory, Chile, and has a resolving power of $R = 48,000$. We obtained 27 spectra with FEROS from our survey and found one additional FEROS spectrum in the ESO archive which was taken on February~22nd, 2012 (see Table A.2). 
 
From these 28 spectra, the RVs were obtained with the Collection of Elemental Routines for Echelle Spectra (CERES) pipeline \citep{Brahm2017}. The RVs were combined into nightly bins, and three strong outliers were omitted.  ($>150\,\mathrm{m\,s}^{-1}$ away from the bulk of the data). 
They are the measurements obtained at epochs 2455879.83, 2458019.89, 
and at epoch 2458094.79 (see \autoref{tab:FEROS_RVdata}).
This left 23 remaining FEROS RVs for our analysis. Except for the outlier RV taken in 2012, all data were taken between July 11th, 2017 and November 27th, 2019, thereby covering a useful temporal baseline of $\sim$869\,d.
After fitting a constant offset of 33629.7 m\,s$^{-1}$ the weighted root-mean-square of the RVs is $wrms_{\mathrm{FEROS}} = 32.4$ m\,s$^{-1}$ with a mean RV uncertainty of $\sigma_{\mathrm{FEROS}}= 4.7$ m\,s$^{-1}$.

To analyze the stellar activity, we determined time series of the emission in the H$_{\alpha}$, Na\,I D1+D2, and Ca~{\sc ii} H\&K lines, by using the continuum normalized flux spectra derived by CERES \citep{Brahm2017}. We went with the approach by \citet{Kuerster2003}, who defined the line index $I$ as a measure of the line emission by 
\begin{equation}
    I = \frac{\langle F_0 \rangle}{0.5 (\langle F_1 \rangle + \langle F_2 \rangle)}, \label{eq:Halpha}
\end{equation}
where $ \langle F_0 \rangle$ is the mean spectral flux of an RV interval centered around the core of the line and $ \langle F_1 \rangle$ and $ \langle F_2 \rangle$ are the mean fluxes in two reference regions on either side of the line, ``bluewards'' and ``redwards'', which are assumed to be constant. 

For the H$_\alpha$ index an RV interval of [$-200$, $+200$] km\,s$^{-1}$, centered on the core of the line, was chosen for the mean spectral flux. 
For the echelle order containing the H$_\alpha$ line, only the ``blueward'' reference region, $ \langle F_1 \rangle$, with an RV interval of [$-1280$, $-800$] km\,s$^{-1}$, could be used, as the order ends shortly redwards of the H$_\alpha$ line.

 \begin{figure}
    \centering
    \includegraphics[width=8.2cm]{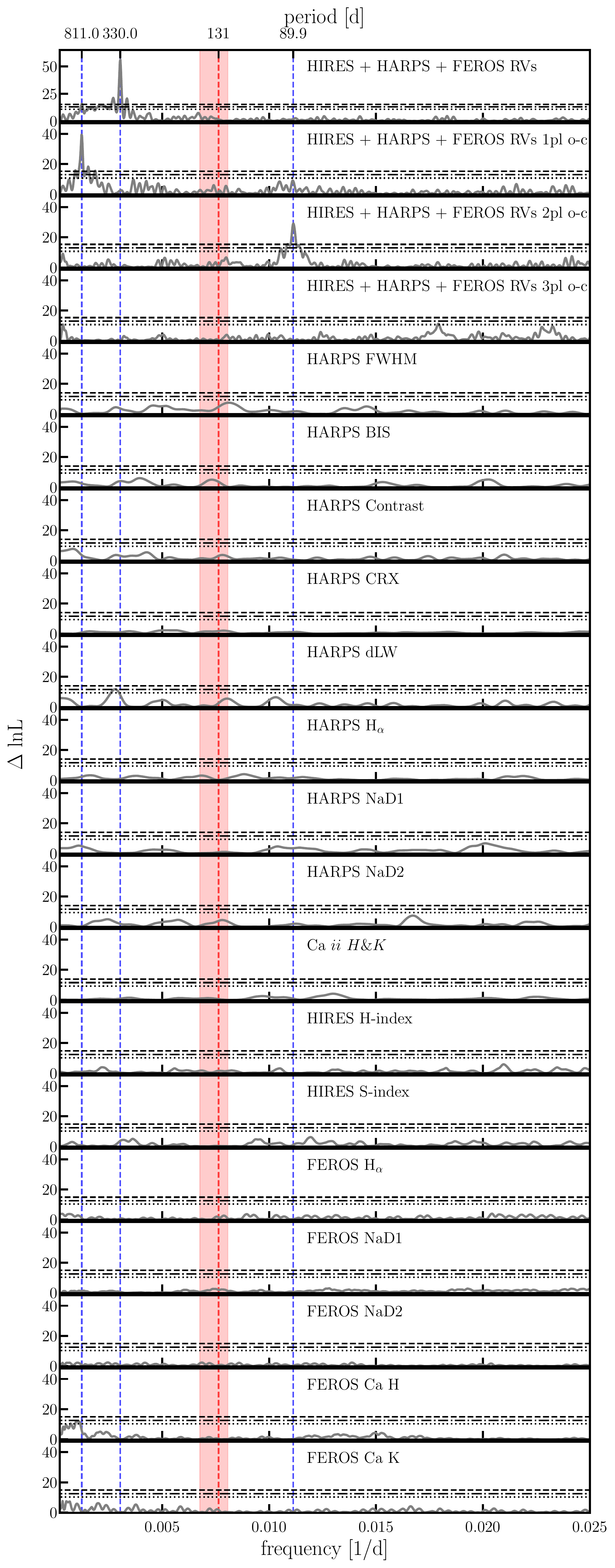}

    \caption{Maximum likelihood periodogram (MLP) of the combined RV data from HARPS, HIRES and FEROS, as well as stellar activity indicators. The horizontal lines correspond to the FAP levels of 10\% (dotted line), 1\% (dot-dashed line), and 0.1\% (dashed line). The orbital periods of the planets are indicated as vertical dashed blue lines and the estimate for the rotation period by \citet{Jofre2015} is shown as a red dashed line, while the red shaded area denotes its uncertainties (see Table \ref{table_st_param}). }
\label{fig:MLP} 
\end{figure}

The RV intervals of the mean central flux $ \langle F_0 \rangle$ for the NaI D1 and NaI D2 line indices are [$-100$, $+100$] km\,s$^{-1}$, centered on the core of the respective line. For the reference bandpasses $ \langle F_1 \rangle$ and  $ \langle F_2 \rangle$, we chose the RV intervals [$-1100$, $-700$] km\,s$^{-1}$ and [$+900$, $+1190$] km\,s$^{-1}$, with the NaI D1 line at a relative RV of zero as reference.

For the Ca\,II H and K indices, we used a flux interval of $\pm 1$ \AA, similar to the intervals used for the HIRES S-index \citep{Butler2017}. For the reference intervals we chose the intervals [$+15$, $+30$] {\AA} and [$-15$, $-25$] {\AA} with the reference lines Ca\,II H and K at zero. For the epochs of 2458127.642 and  2458127.645 it was not possible to calculate these indices due to the low signal-to-noise ratio of the spectra.

\vspace{1cm}

\subsection{HARPS Data} 
\label{Sec3.2}
The High Accuracy Radial velocity Planet Searcher (HARPS) is a precise cross-dispersed échelle spectrograph installed at the ESO 3.6 meter telescope at La Silla Observatory, Chile. It has a spectral resolution of $R = 115,000$ and reaches an RV measurement precision of $\sim$ 1 m\,s$^{-1}$ \citep{Mayor2003}. We obtained 88 high signal-to-noise spectra for HD\,33142 within our HARPS survey, and found 10 more HARPS spectra for this star in the ESO archive. Thus, we have a total of 98 spectra, taken from Aug.~9th, 2016 to Sept.~28th, 2019, covering a temporal baseline of 1145.0\,d. From these spectra we determined high-precision RV data and spectral activity data using the SpEctrum Radial Velocity AnaLyser (\texttt{serval}) \citep{Zechmeister2018} and the HARPS-DRS (Data Reduction Software) pipeline offered by ESO \citep{Mayor2003}. 
After fitting a constant offset of $-25.5\,\mathrm{m\,s}^{-1}$ the weighted root-mean-square of the RVs is $\textnormal{wrms}_{\mathrm{HARPS}} = 20.00$ m\,s$^{-1}$ with a mean RV uncertainty of $\sigma_{\mathrm{HARPS}}= 0.59$ m\,s$^{-1}$.

From \texttt{serval} the chromatic index (CRX, a measure of the wavelength dependence of the RV signal) and the differential line width (dLW, a measure of the width of the average absorption line) were obtained 
\citep[see][for more on these quantities]{Zechmeister2018}, as well as 
activity indices for H$\alpha$, Na\,I D1 and Na\,I D2.  From HARPS-DRS the parameters of a cross-correlation with a line-list template are obtained, namely, the full width at half maximum (FWHM), bisector inverse slope (BIS), and Contrast of the cross-correlation function \citep[for more on this method see][]{1995IAUS..167..221Q,2002A&A...388..632P}. We also inspect the HARPS Ca~{\sc ii}~H\&K chromospheric activity time series, following the approach developed by \citet[][see their paper for more details]{Perdelwitz2021}.

\subsection{HIRES Data}
\label{Sec3.3}
The High Resolution Echelle Spectrometer (HIRES) is a grating cross-dispersed échelle spectrograph mounted on the KECK 10 meter telescope on Mauna Kea, Hawaii \citep{Vogt1994}. 
It has a spectral resolving power of $R = 25,000 - 85,000$, depending on the chosen slit and can reach a precision of $\sim 3\,\mathrm{ms}^{-1}$ when the highest resolution is used \citep{Butler1996}.

We use 40 archival HIRES spectra taken from October 2007 to March 2020 corresponding to a temporal baseline of $\sim$4500\,d.
To derive RVs, and stellar activity index data (S-index and H-index) we used the same data  reduction pipeline as used by \citet{Butler2017}.
After fitting a constant offset of $-9.17\,\mathrm{m\,s}^{-1}$ the weighted root-mean-square of the RVs is $\textnormal{wrms}_{\mathrm{HIRES}} = 22.6$ m\,s$^{-1}$ with a mean RV uncertainty of $\sigma_{\mathrm{HIRES}}= 1.7$ m\,s$^{-1}$.\looseness=-4

\begin{figure*}
    \centering
    \includegraphics[width=18cm]{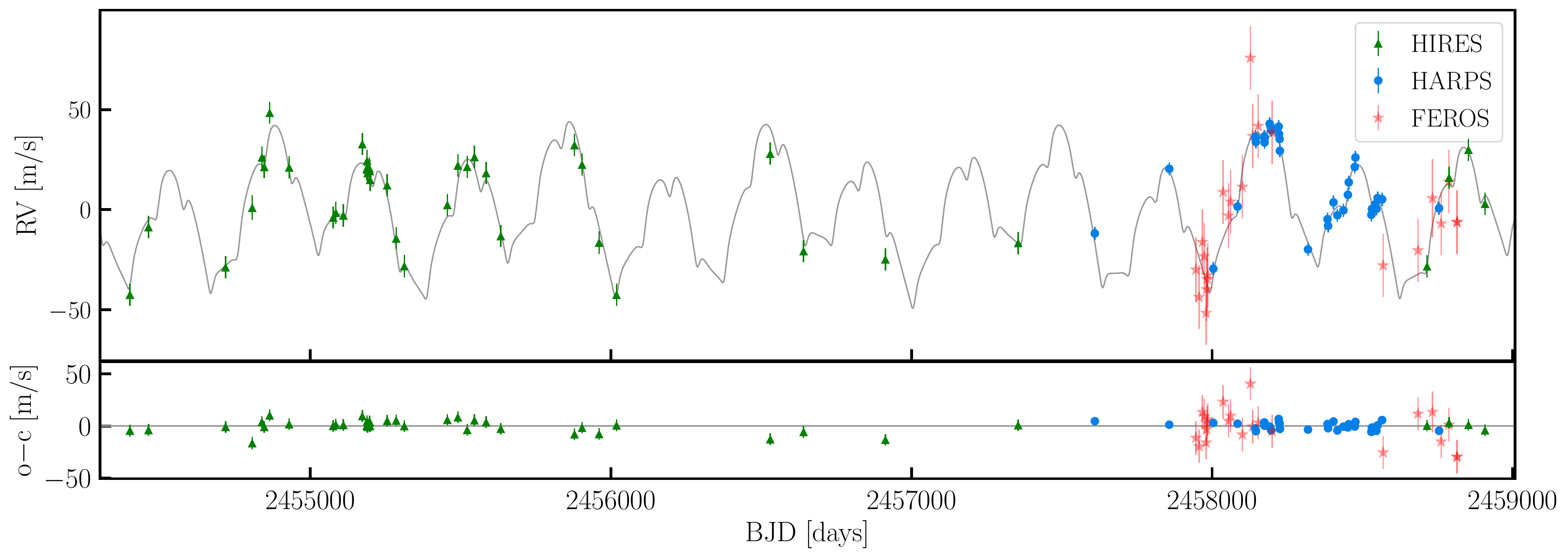}
    \includegraphics[width=18cm]{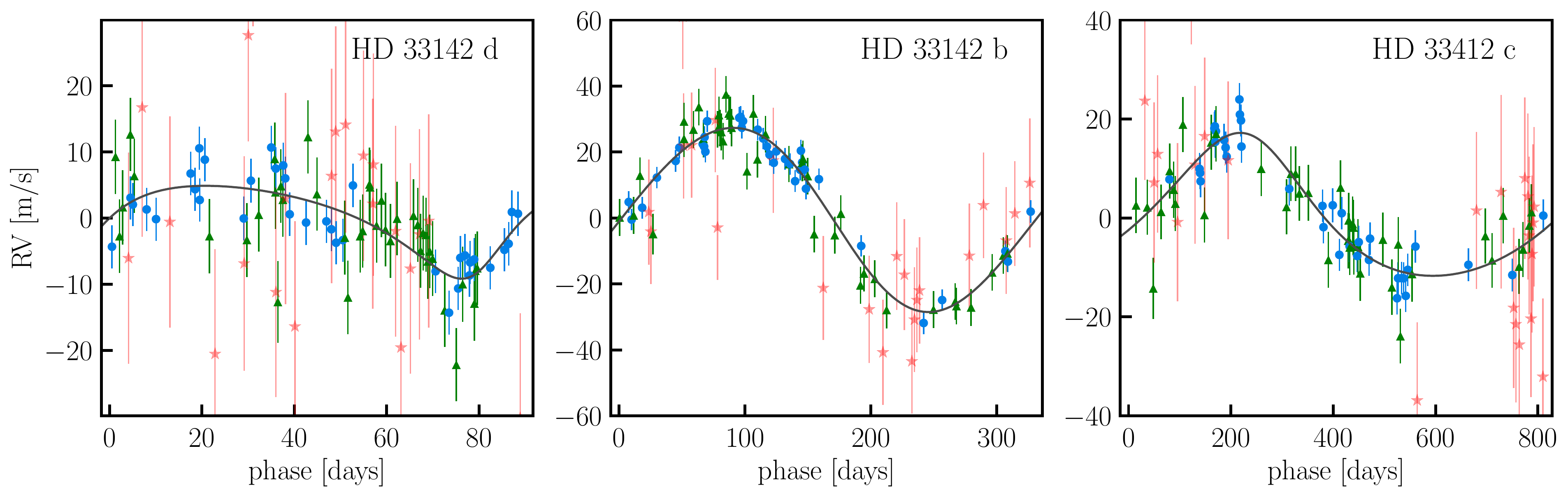}
 
    \caption{Top main panel shows the RV measurement time series from HIRES (green triangles), HARPS (blue circles), and FEROS (red stars) and the best three-planet model for HD\,33142 (gray line, \autoref{table:best_fit}). Top small panel shows the data residuals after applying the best three-planet Keplerian model. The data uncertainties include the RV jitter. The bottom panels show phase-folded representation of the planetary signals. In each case, the orbits of the other two planets were subtracted.}
\label{fig:RV_curve} 
\end{figure*}

\section{Analysis and Results}
\label{Sec4}

\subsection{Periodogram Analysis}
\label{Sec4.1}

We inspect the available radial velocity and stellar activity data using a maximum likelihood periodogram (MLP) following the methodology used by \citet{Zechmeister2019} and \citet{Trifonov2021a}.
The MLP algorithm is a more advanced implementation of the generalized Lomb-Scargle periodogram \citep{Zechmeister2009}, which optimizes log-likelihood $\ln\mathcal{L}$, for each scanned frequency by allowing  simultaneous fitting of multiple telescope data sets, each with separate additive offset and a jitter term \citep[see also,][]{Baluev2009}. We adopted significance thresholds of the likelihood improvements with respect to a null model with  an amplitude equal to 0, which corresponds to false-alarm probabilities (FAPs) of 10\%, 1\%, and 0.1\%. For the significance threshold, a FAP value of $< 0.1\%$ was chosen.

In \autoref{fig:MLP}, the MLPs of the combined RV data from HARPS, HIRES and FEROS (upper four panels), as well as those of several activity indicators (lower fifteen panels) are shown. The most probable stellar rotation period by \citet{Jofre2015},  $P_\mathrm{rot}/\sin{i}=132 \pm 18\,\mathrm{d}$ is indicated as a red dashed line and its uncertainty is shown as a red shaded region.  False alarm probability (FAP) levels of 10\%, 1\%, and 0.1\% are indicated as dotted, dot-dashed and dashed lines, respectively. 
The first four panels show periodograms of the residuals from a flat model (only individual RV offsets for each dataset fitted) and for a single-, two- and three-planet Keplerian fit, respectively. The flat model residuals show a significant peak at a period of $\sim 330$\,d, which is in good agreement with \citet{Johnson2011} who announced HD\,33142\,b with an orbital period of $P_\mathrm{b} = 326.6$\,d. 
After subtracting a single-planet Keplerian fit with a planetary period of $P_\mathrm{b} = 329.1$\,d, the second panel shows a strong power at a period of $\sim 805$\,d which is in good agreement with \citet{Bryan2016} and \citet{Luhn2019}.
The third panel shows the residuals after subtracting a combined two-planet Keplerian fit with planetary periods $P_\mathrm{b} = 330.2$\,d and $P_\mathrm{c} = 811.1$\,d. A strong power at a period of $\sim 90$\,d can be seen. Performing a combined three-planet Keplerian fit yields planetary periods of $P_\mathrm{b} = 330.5$\,d, $P_\mathrm{c} = 807.5$\,d and $P_\mathrm{d} = 89.9$\,d, indicated in the panels as blue dashed lines. None of these periods coincides at face value or within their 1-$\sigma $ errors with the possible rotation periods listed in Table \ref{table_st_param}. The fourth panel shows the periodogram of these RV residuals after the three-planet Keplerian fit. No further significant peaks are visible.

To analyze whether any of these periodicities are due to stellar activity and not induced by a planet, a periodogram analysis of the activity data is performed. Starting downwards from the fifth panel (counted from the top), 
Fig.~\ref{fig:MLP} shows the MLPs for different activity indicators as labeled in the panels. No significant peaks in the activity data MLPs can be seen near the detected periodicities, thus confirming the planetary nature of these signals.  The only noteworthy periodogram peak near one of the detected periodicities is seen in the differential line width dLW indicator (ninth panel from the top), namely around $1\,\mathrm{yr}$. It is wide enough to also encompass the $330\,\mathrm{d}$ period value, but does not peak there.  Its peak has a FAP of a few percent and is therefore not significant.\looseness=-4

\begin{table*}[t!]

 \caption{Comparison of fit quality between different models through $\ln\mathcal{L}$, BIC, and the Bayesian log-evidence $\ln Z$, with respect to the Null-model. For eccentric orbits $e$ and $\omega$ are free fit parameters and for circular orbits $e=0$ and $\omega$ is undefined.
} 
\begin{adjustwidth}{-3.0cm}{}

\centering        

\begin{tabular}{l l r r r r r r r r}     
\hline\hline  \noalign{\vskip 0.5mm}        
  Model  &Planets & &max. $\ln{\mathcal{L}}$   &  BIC &  $\ln Z$ && $|\Delta$ max. $\ln\mathcal{L}|$ & $|\Delta$BIC$|$ & $|\Delta\ln Z|$ \\ 
\hline    \noalign{\vskip 0.5mm}                   
  Null-model & No planet   && $-$447.23   & 882.46  & $-$493.69 & & \dots  & \dots & \dots \\ 
  
   circular  & Planet b    && $-$390.35   & 821.97  & $-$450.99 & &  56.88 & 60.49 & 42.70\\ 
   
  eccentric  & Planet b    && $-$388.48   & 827.40  & $-$451.30 & &  58.75 & 55.06 & 42.39 \\ 
  
  circular   &  Planets b and c  && $-$348.23  &  751.47 & $-$420.61 & & 99.00  & 130.99  & 73.08 \\
  
 eccentric & Planets b and c     && $-$340.75  &  754.86 & $-$432.35 & & 106.48 &  127.60 & 60.34 \\ 
 
  circular &  Planets b, c and d && $-$316.49  &  701.76 & $-$376.48 & & 130.74 & \textbf{180.70} & 117.21 \\ 
  
 eccentric &  Planets b, c and d &&  $-$312.25 &  720.80 & $-$371.34 & &\textbf{134.98}  & 161.66 & \textbf{122.35} \\ 
 
 

  
  
  
   
    

\hline\hline \noalign{\vskip 0.5mm}   

\end{tabular}

\end{adjustwidth}
\label{table:comp} 

\end{table*}

\subsection{Orbital Analysis}
\label{Sec4.2}

To model the planetary orbits from the available precise Doppler data, we adopt the 
{\sc Exo-Striker}\footnote{\url{https://github.com/3fon3fonov/exostriker}} \citep{ES2019} exoplanet toolbox.
We use a Nelder-Mead (Simplex) algorithm
\citep[][]{NelderMead1965} in order to optimize the likelihood function (in the form $-\ln \mathcal{L}$) and return the best-fit parameters of a Keplerian model. Besides the RV model parameters semi-amplitudes $K$, periods $P$, 
eccentricities $e$, arguments of periastron $\omega$, and mean anomalies $M$ defined at the first observational epoch for each star, 
{\sc Exo-Striker} works in Jacobi frame, which is a natural coordinate system for  orbital parameters extracted from RV signal consistent with multiple planet systems.
Additionally, in our models, we vary the RV offset and RV jitter for each individual dataset. Following \citet{Baluev2009} the jitter parameter appears as an additional RV noise term added in quadrature to the nominal RV uncertainties and subsuming any instrumental or stellar effects (such as activity) not considered by the formal RV errors. We adopt a purely Keplerian model, in which the planetary inclinations $i$ and the difference of the orbital ascending nodes $\Delta \Omega$ cannot be accessed, and assume a coplanar and edge-on configuration of the system.

We estimate the parameter posterior distributions using an affine-invariant ensemble Markov chain Monte Carlo (MCMC) sampler
\citep{2010CAMCS...5...65G} from the \texttt{emcee} package \citep{emcee}, integrated within the {\sc Exo-Striker} program.
We adopted non-informative flat priors and used the best-fit parameters returned by the Simplex minimization as the MCMC chain starting point for the exploration of the parameter space. We ran 120 independent walkers in parallel in a standard
scheme of 1,000 burn-in MCMC steps, which we discarded from the analysis,
followed by 5,000 MCMC steps from which we constructed the parameter
posterior distribution. At the end of each MCMC run, we evaluated the
sampler acceptance fraction to ensure that the MCMC chains have converged
\citep[which should be between 0.2 and 0.5, see][]{emcee}.

Finally, the MCMC posteriors were used to assess prior estimates, which are used 
for further posterior analysis using the nested sampling (NS) technique \citep{Skilling2004}. For the purpose, the {\sc Exo-Striker} adopts the {\tt dynesty} sampler \citep{Speagle2020}.
Our NS setup consists of 100 live points per fitted parameter, sampled via random walk dynamic nested sampling.
We found the NS technique particularly useful and somewhat superior over the MCMC, because it can efficiently study a large parameter space within the adopted prior space and calculate the Bayesian log-evidence $\ln Z$, which is an important diagnostic when comparing competitive models \citep[e.g.,][among many more recent exoplanet studies]{Luque2019, Espinoza2019, Stock2020, Trifonov2021a,Trifonov2021b}.\looseness=-4

\begin{table*}[ht]
    \centering
    \caption{Best-fit parameters, posterior estimates for the three-planet Keplerian fits valid for the time of the first observation, BJD = 2454400.032. The best-fit parameters are obtained from a Simplex MLE scheme, and the median parameter values and uncertainties come from NS analysis by using 68.3\% of the confidence level of the dynamically stable posteriors (667 stable samples out of 1000 randomly chosen samples from the total NS size of 29\,828 unique samples).
    The adopted priors are listed in the right-most columns and their meanings are $\mathcal{U}$ -- Uniform, 
    and $\mathcal{J}$ -- Jeffrey's (log-uniform) priors.
    }
    
    \label{table:best_fit}
      \centering
  
     \begin{adjustwidth}{-0.0cm}{}
     \resizebox{0.73\textheight}{!}
     {\begin{minipage}{1.1\textwidth}

    \centering

    \begin{tabular}{lrrr|rrr|rrrrrr}     

    \hline\hline  \noalign{\vskip 0.7mm}
\makebox[0.1\textwidth][l]{\hspace{40 mm}  Max. $-\ln\mathcal{L}$   \hspace{30 mm}  Median and $1\sigma$ (stable)  \hspace{40 mm} Adopted priors  \hspace{10 mm}  } \\

    \hline   \noalign{\vskip 0.7mm}
    Parameter \hspace{0.0 mm}& Planet d & Planet b & Planet c  &  Planet d & Planet b & Planet c  & Planet d & Planet b & Planet c\\
    \hline \noalign{\vskip 0.7mm}

        $K$ [m\,s$^{-1}$]             &   7.01  &    27.95 &    14.47    
                                      &   7.1$_{-0.9}^{+0.8}$  &    28.1$_{-1.1}^{+1.1}$ &    14.7$_{-0.9}^{+0.9}$    
                                      &  $\mathcal{U}$(0.0,20.0) & $\mathcal{U}$(0.0,40.0) & $\mathcal{U}$(0.0,40.0)   \\ \noalign{\vskip 0.9mm}

        $P$ [d]                     &    89.89 &   329.72 &   811.50       
                                      &    89.9$_{-0.1}^{+0.1}$ &   330.0$_{-0.4}^{+0.4}$ &   810.2$_{-4.2}^{+3.8}$     
                                      &  $\mathcal{U}$(89.0,91.0) & $\mathcal{U}$(320.0,340.0) & $\mathcal{U}$(800.0,820.0)   \\ \noalign{\vskip 0.9mm}
                                      
        $e$                           &     0.352 &     0.052 &     0.197        
                                      &     0.191$_{-0.128}^{+0.140}$ &     0.049$_{-0.030}^{+0.032}$ &     0.081$_{-0.047}^{+0.055}$      
                                      &  $\mathcal{U}$(0.0,0.5) & $\mathcal{U}$(0.0,0.5) & $\mathcal{U}$(0.0,0.5)   \\ \noalign{\vskip 0.9mm}

        $\omega$ [deg]                &   209.8 &   115.8 &   16.3     
                                      &   231$_{-29}^{+40}$ &   129$_{-38}^{+46}$ &   52$_{-57}^{+71}$
                                      &  $\mathcal{U}$(0.0,360.0) & $\mathcal{U}$(0.0,360.0) & $\mathcal{U}$(0.0,360.0)   \\ \noalign{\vskip 0.9mm}

        $M_{\rm 0}$ [deg]             &     357.5  &    95.8  &   93.3      
                                      &    346$_{-37}^{+26}$ &      83$_{-45}^{+39}$ &   58$_{-72}^{+56}$     
                                      &  $\mathcal{U}$(0.0,360.0) & $\mathcal{U}$(0.0,360.0) & $\mathcal{U}$(0.0,360.0)   \\ \noalign{\vskip 0.9mm}

        $a$ [au]                      &     0.452  &     1.075  &     1.957        
                                      &     0.452$_{-0.003}^{+0.003}$ &     1.074$_{-0.007}^{+0.007}$ &     1.955$_{-0.012}^{+0.016}$       
                                      &  (derived) & (derived) & (derived)   \\ \noalign{\vskip 0.9mm}

        $m \sin i$ [$M_{\rm Jup}$]    &     0.20  &     1.261 &     0.885      
                                      &     0.20$_{-0.03}^{+0.02}$ &     1.26$_{-0.05}^{+0.05}$ &     0.89$_{-0.05}^{+0.06}$   
                                      &  (derived) & (derived) & (derived)  \\ \noalign{\vskip 0.9mm}

        RV$_{\rm off, HIRES}$ [m\,s$^{-1}$]& &  -14.9  & \hspace{1.0 mm}& \hspace{1.0 mm}
                                           &   -15.7$_{-1.4}^{+1.1}$ & \hspace{1.0 mm}& \hspace{1.0 mm}
                                           &   $\mathcal{U}$(-100.0,100.0) &    \\ \noalign{\vskip 0.9mm}
                                           
        RV$_{\rm off, HARPS}$ [m\,s$^{-1}$]& &  -40.7  & \hspace{1.0 mm}& \hspace{1.0 mm}
                                           &   -40.8$_{-0.9}^{+0.9}$ & \hspace{1.0 mm}& \hspace{1.0 mm} 
                                           &  $\mathcal{U}$(-100.0,100.0) &    \\ \noalign{\vskip 0.9mm}

        RV$_{\rm off, FEROS}$ [m\,s$^{-1}$]& & 33634.2 & \hspace{1.0 mm}& \hspace{1.0 mm}
                                           &   33633.9$_{-3.8}^{+3.4}$ & \hspace{1.0 mm}& \hspace{1.0 mm} 
                                           &   $\mathcal{U}$(33600.0,33800.0) &    \\ \noalign{\vskip 0.9mm}

        RV$_{\rm jit, HIRES}$ [m\,s$^{-1}$]& &   5.3  & \hspace{1.0 mm}& \hspace{1.0 mm}
                                           &   6.8$_{-0.9}^{+0.9}$ & \hspace{1.0 mm}& \hspace{1.0 mm} 
                                           &   $\mathcal{J}$(0.1,50.0) &    \\ \noalign{\vskip 0.9mm}

        RV$_{\rm jit, HARPS}$ [m\,s$^{-1}$]& &   3.2 & \hspace{1.0 mm}& \hspace{1.0 mm}
                                           &   3.4$_{-0.5}^{+0.7}$ & \hspace{1.0 mm}& \hspace{1.0 mm}
                                           &   $\mathcal{J}$(0.1,50.0) &    \\ \noalign{\vskip 0.9mm}

        RV$_{\rm jit, FEROS}$ [m\,s$^{-1}$]& &  15.3 & \hspace{1.0 mm}& \hspace{1.0 mm}
                                           &   16.8$_{-2.5}^{+3.6}$ & \hspace{1.0 mm}& \hspace{1.0 mm}
                                           &   $\mathcal{J}$(0.1,50.0) &    \\ \noalign{\vskip 0.9mm}

        \noalign{\vskip 0.7mm}
    \hline \noalign{\vskip 5.7mm}

    \end{tabular}

     \end{minipage}}
     \end{adjustwidth}


    \end{table*}

To further constrain the orbital configuration of the system, we compare the three-planet model against the two-planet, single-planet, and null-hypothesis model with no planets. We also test eccentric configurations of the system against circular configurations. For the latter ones, the eccentricity is forced to be zero and the argument of periastron is undefined. For our comparison, we use three quality-of-fit quantities: the log-likelihood $\ln\mathcal{L}$, the Bayesian Information Criterion (BIC), and the Bayesian log-evidence $\Delta\ln Z$. 
The BIC attempts to resolve the problem of ``overfitting'' (as a decrease in likelihood can be achieved through adding parameters) by introducing a penalty term for the number of parameters in the model. It is defined by
\begin{equation}
    \mathrm{BIC} = k \ln n - 2 \ln \mathcal{L},
\end{equation}
where $k$ is the number of parameters in the model, $n$ the number of data points and $\mathcal{L}$ the maximized likelihood of the model. When comparing a simpler model to a more complex one, we compute $\Delta\ln{\mathcal{L}} = |\ln{\mathcal{L}}_{\rm simple}| - |\ln{\mathcal{L}}_{\rm complex}|$ and $\Delta\mathrm{BIC} = \mathrm{BIC}_{\rm simple} - \mathrm{BIC}_{\rm complpex}$. A strong evidence for a more complex model to be preferred over the
simpler one is $\Delta\ln{\mathcal{L}}>7$ (strictly speaking only for models with the same degree of freedom)\footnote{In this case the relative probability is $R = e^{-\Delta \ln{\mathcal{L}}}\approx 0.001$, corresponding to $\sim 0.1\% $ FAP necessary for a significant detection \citep{Anglada-Escude2016}} and 
$\Delta\mathrm{BIC}>10$ \citep{Kass1995}. For the Bayesian log-evidence comparison we followed \citet{Trotta2008}, who considered two models indistinguishable if their Bayesian log-evidence difference satisfies $\Delta \ln Z \lesssim 2$, a model moderately favored over another if $\Delta\ln Z$ $>$ 2, and strongly favored if $\Delta \ln Z > 5$.


In Table \ref{table:comp}, the $\ln{\mathcal{L}}$, BIC, $\ln Z$, values are listed along with their differences with respect to the ``Null''-model, whose only fit parameters are the RV-offsets and RV-jitter for each of the three individual data sets. 
Due to their high $\Delta \ln{\mathcal{L}}$, $\Delta \mathrm{BIC}$, and $\Delta \ln Z$ values, it is very suggestive that in all cases the planet models on circular or eccentric orbits present a significant improvement over the base model with no planet. Comparing the one-~, two-~, and three-planet models between each other, 
we observe significant improvement for every added planetary companion. In 
terms of $\Delta \ln{\mathcal{L}}$ the eccentric fits have only marginal improvement with respect to the circular orbits, while $\Delta \mathrm{BIC}$ practically penalizes the eccentric solutions.
For instance, by comparing the circular and eccentric three-planet configurations (planets b, c, and d)
we conclude that the $\Delta \ln{\mathcal{L}}$ improvement of the more complex model is not significant, whereas $\Delta \mathrm{BIC}$ clearly favors the simpler, circular fit. The small $\Delta \ln{\mathcal{L}}$ difference between the eccentric and the circular configuration is likely due to all three planetary eccentricities being small and poorly constrained. The fact that $\Delta \mathrm{BIC}$ favors the simpler model can be explained as for an eccentric three-planet configuration, six more fit parameters are needed than for the circular configuration, and since $\Delta \ln{\mathcal{L}}$ for both configurations is similar, BIC favors the circular configuration due to the penalty term. 

Finally, $\Delta \ln Z$ suggests no difference between one-planet circular and eccentric fit, but favors the eccentric solutions for the two- and three-planet fits (although admittedly close to the $\Delta \ln Z$ significance threshold). We achieve the highest $\Delta \ln Z$ for the three-planet eccentric fit, with $\Delta \ln Z$ = 122.35 with respect to the ``Null''-model, and $\Delta \ln Z$ = 5.14 with respect to the three-planet circular model.
Therefore, for our orbital evolution analysis presented in Section \ref{Sec4.4} we choose to adopt the three-planet eccentric configuration. A perfectly circular orbit is unlikely to be found in nature, hence a small eccentricity is more realistic.

The best-fit parameters together with their NS estimated errors for the three-planet model for HD\,33142 are listed in Table \ref{table:best_fit}. We conclude that the HD\,33142 planetary system comprises two outer Jupiter-like planets and one inner Saturn-like planet. The final NS  parameter posterior distributions from this work are shown in the Appendix in Fig.~\ref{fig:CP}.

\begin{figure*}
    \includegraphics[width=17.32cm]{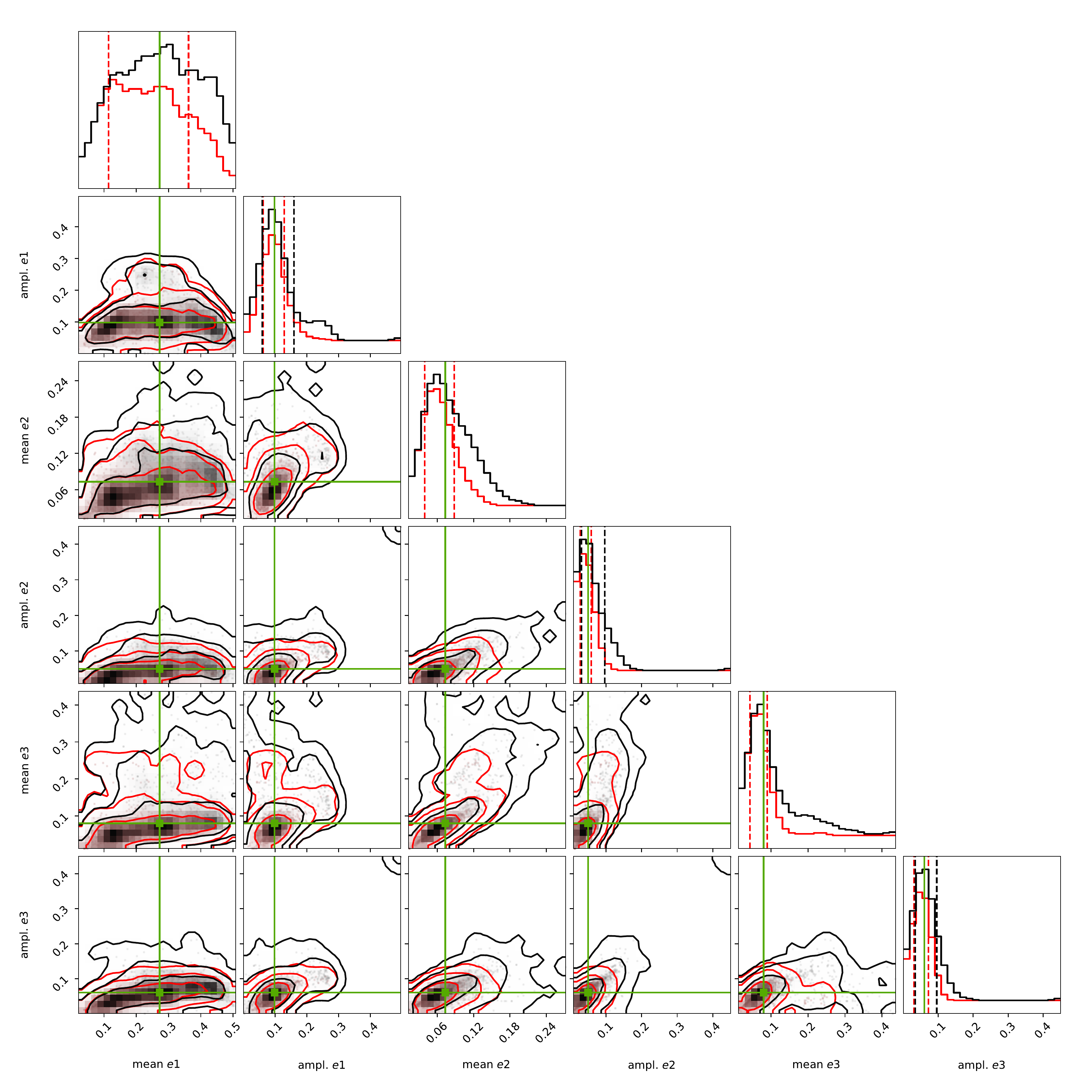}  
    \caption{Distribution of 1\,000 randomly sampled NS posteriors integrated for 1\,Myr. The black, two-dimensional contours indicate 1-, 2- and 3-$\sigma$ confidence intervals of the whole posterior distribution, and the red contours indicate 1-, 2- and 3-$\sigma$ confidence intervals for the stable sub-sample. Only the mean planetary eccentricities and eccentricity amplitudes for all three planets are shown, which are decisive for the system stability and dynamical properties. The median values of the total posterior distribution are marked with green lines.
    }
    \label{Fig:dyn_mcmc}
\end{figure*}

The time series of the 98 radial velocity measurements of HD\,33142 are shown in Figure \ref{fig:RV_curve}, along with the best three-planet Keplerian fit to the data. Different colors correspond to different instruments, as indicated at the top right in the upper panel.
The lower part of the upper panel shows the residual RV data after subtraction of the best three-planet Keplerian fit. The bottom panels show the phase-folded representation of the planetary signals from HD\,33142\,d, HD\,33142\,b and HD\,33142\,c. 
The HARPS and HIRES RVs are in good agreement with the best-fit Keplerian model, but the FEROS RVs deviate more. 
This can also be seen in the RV scatter around the best-fit model: the RV scatter of HARPS, HIRES, and FEROS is $\textnormal{rms}_{\mathrm{HARPS}} = 3.3$\,m\,s$^{-1}$, $\textnormal{rms}_{\mathrm{HIRES}} = 5.8$\,m\,s$^{-1}$, and 
$\textnormal{rms}_{\mathrm{FEROS}} = 15.9$\,m\,s$^{-1}$, respectively.

\subsection{Dynamical Analysis}
\label{Sec4.4}

The long-term dynamical analysis of the HD\,33142 system
was carried out using the symplectic N-body integrator {\sc SyMBA} \citep{Duncan1998}, implemented in the {\sc Exo-Striker}. The {\sc SyMBA} algorithm is designed to accurately compute
close planetary encounters if such occur during the orbital evolution \citep[see,][for details]{Duncan1998}. Else, in the case of well-separated orbits, {\sc SyMBA} is as efficient as other well-established N-body algorithms such as the Wisdom-Holman mapping method \citep{Wisdom1991}. In addition, our version of {\sc SyMBA} is customized to work in the Jacobi coordinate system \citep[e.g.,][]{Lee2003}, the same as our output orbital parameters. We assume a coplanar prograde configuration, meaning that all planets are moving in the same  
direction.

\begin{figure*}[t]
     \centering
     \includegraphics[width=8.85cm]{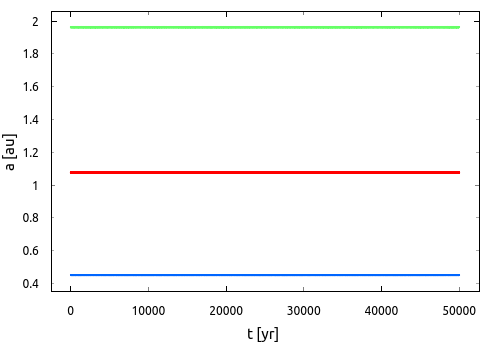}
     \includegraphics[width=8.85cm]{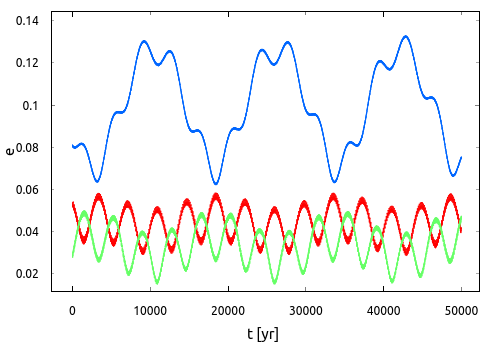} \\
     \includegraphics[width=5.9cm]{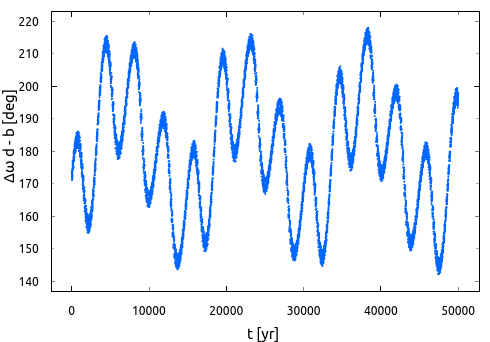}
     \includegraphics[width=5.9cm]{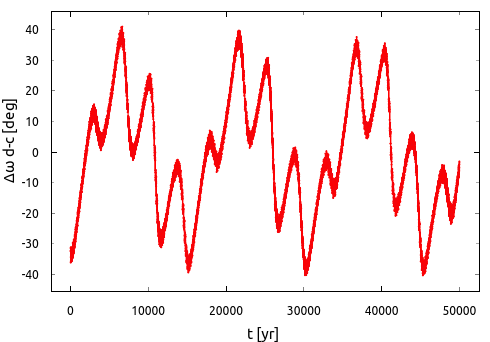}
     \includegraphics[width=5.9cm]{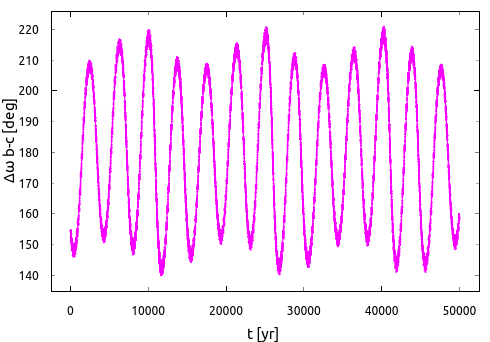} \\

     \caption{Top row: The evolution of the semi-major axes $a$ and eccentricities $e$ for HD\,33142\,d (blue), HD\,33142\,b (red), and HD\,33142\,c (green) of a representative stable fit from the posterior distribution. Middle row: The evolution of apsidal angles $\Delta\omega_{\rm d-b}$,
     $\Delta\omega_{\rm d-c}$, and $\Delta\omega_{\rm b-c}$, which are librating in anti-aligned, aligned, and anti-aligned geometry, respectively.
     For purposes of clear representation, the plot only shows a 50\,000 yr excerpt.}
     \label{fig:ecc}
 \end{figure*}

To assess the system's dynamical properties and analyze how the starting configuration impacts the stability, we chose 1\,000 randomly selected samples from the NS posterior distribution (a total of, 29\,828 unique samples) as starting points and integrated them for 1\,Myr, with a 0.5\,d time step.
Our preliminary numerical integrations suggest that the system's instability usually occurs in short timescales of less than 100\,000\,yr; thus, the timescale of  1\,Myr is sufficient for testing the overall dynamics of the system given the estimated posterior parameter space. 
Following \cite{Trifonov2020}, we adopt the condition that the planetary semi-major axes must not deviate at any given time of the orbital evolution by more than 10\% from the adopted initial values of the NS sample, while the eccentricities are not allowed to lead to crossing orbits.
Otherwise, the sample is rejected as unstable. This ensures that the system remains regular and well separated.\looseness=-4

We also use the N-body simulations to search for possible mean
motion resonances (MMR) between the planets by calculating the time evolution of the MMR angles, which we define in the following. In an $n':n$ MMR between an inner planet $i$ and an outer planet $o$, the corresponding resonance angles are given by:
\begin{equation}
    \phi_{mnn'} = n\lambda_i - n'\lambda_o + (m-n)\varpi_i - (m-n')\varpi_o,
    \label{eq:resonance_angles}
\end{equation}
where $\lambda = M + \varpi $ is the mean longitude with $M$ being the mean anomaly, $\varpi = \omega + \Omega$ is the longitude of periastron with $\omega$ being the argument of periastron and $\Omega$ being the longitude of the ascending node (which in our case is undefined and set to $\Omega$=0$^\circ$), and $m$ is an integer satisfying $n \leq m \leq n'$ \citep{Mardling2013}.
To determine whether the HD\,33142 system is in an MMR configuration, we start with calculating the orbital period ratios between the planets.
The approximate period ratios from the best-fit for planets b and d, c and b, and c and d are, 
respectively, $P_{\rm rat. b,d} \sim 3.7, P_{\rm rat. c,b}  \sim 2.5$, and 
$P_{\rm rat. c,d}  \sim 9.0$, corresponding to possible MMR resonances of 11:3, 5:2, and 9:1.
Generally, these are high-order MMRs with rather low probability of occurrence. Nevertheless, we test whether these period ratios are really indicative of an MMR  behavior, and we analyze their time evolution.

Additionally, we monitor for libration of the secular apsidal angle $\Delta\omega$ for each pair of planets, which is defined as:
\begin{equation}
\Delta\omega_{b-c}=\varpi_b-\varpi_c,~~~ \Delta\omega_{d-b}=\varpi_d-\varpi_b,~~~ \Delta\omega_{d-c}=\varpi_d-\varpi_c. \\
\end{equation}
Libration of any of the $\Delta\omega$ without MMR would indicate that the dynamics of the system is dominated by secular apsidal alignment.

\begin{figure*}[t]
     \centering
 
      \includegraphics[width=5.9cm]{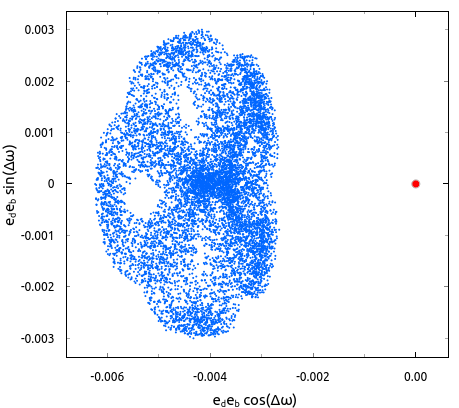}
     \includegraphics[width=5.9cm]{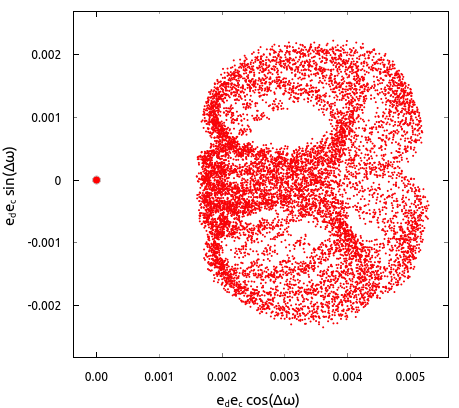}   
       \includegraphics[width=5.9cm]{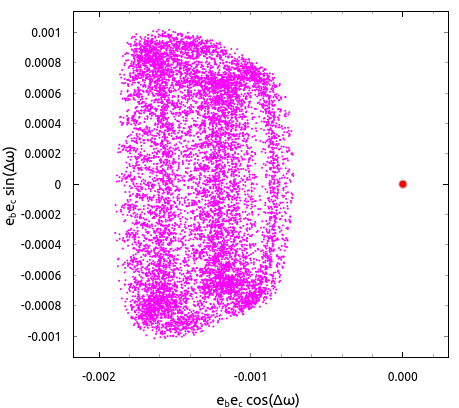} \\

     \caption{Same evolution as in \autoref{fig:ecc} but for the trajectory evolution of HD\,33142 d,\,b \& c, as a function of the product of their eccentricity and sine/cosine of their mutual $\Delta\omega$. In all cases, these variables exhibit libration around a fixed point (red dot, centered at [0,0]).     
     }
     \label{fig:ecc2}
 \end{figure*}

From the 1\,000 randomly selected samples, we find that 66.7\% are stable over the considered 1\,Myr time. For the remaining 33.3\% samples, we find a median survival time of only t$_{\rm max.}$ $\approx$ 8500\,yr.
Fig.~\ref{Fig:dyn_mcmc} shows the distribution of the
 mean planetary eccentricities and end-to-end (minimum to maximum) eccentricity amplitudes of the integrated 1\,000 samples. These dynamical parameters are determinants for the system's dynamical properties and long-term stability. For the stable samples, the mean is taken over the 1\,Myr period, and for the unstable samples, the mean is taken over the time interval before they become unstable. The green lines indicate the position of the median values from the full posterior distribution. The black two-dimensional contours indicate 1-, 2- and 3-$\sigma$ confidence intervals of the whole posterior distribution and the red contours indicate 1-, 2- and 3-$\sigma$ confidence intervals for the stable sub-sample. These simulations suggest that larger posterior planetary eccentricities 
 generally lead to unstable orbits in short time scales. For the stable samples, no MMR configurations were uncovered.

We integrate a few dozen three-planet configurations of HD\,33142 constructed from the stable posterior samples for 100\,Myr, and we find these stable within this baseline. This gives us the confidence that at least a very large fraction of the 1\,Myr stable posterior sample is long-term stable. Our overall posterior dynamical analysis suggests that the MMR angles exhibit only circulation between 0$^\circ$ and 360$^\circ$.  Therefore, no resonance configuration between the planets is indicated.
However, for a moderate fraction of the posteriors we observe a clear libration of the apsidal angle of the two Jovian mass-planets HD\,33142 b \& c, $\Delta\omega_{b,c}$ $\sim$ 0$^\circ$, or even between all three planets, where in most cases  $\Delta\omega_{b,c}$ librates around 180$^\circ$,  $\Delta\omega_{d,b}$ around 180$^\circ$, and $\Delta\omega_{d,c}$ around 0$^\circ$. 

 Fig.~\ref{fig:ecc} shows a 50\,000\,yr excerpt of the evolution of a stable sample with librating $\Delta\omega_{b,c}$, $\Delta\omega_{d,b}$, and $\Delta\omega_{d,c}$. In the example configuration, the orbital eccentricities exhibit moderate oscillations, whereas the semi-major axes remain constant. The eccentricity of the innermost planet d (blue) varies between 0.06 and 0.13, that of planet b (red) between 0.03 and 0.06, and that of planet c (green) between 0.02 and 0.05. As can be seen, the eccentricities of planets b and c largely vary in anti-phase. These two planets exchange energy and momentum, and thus eccentricity. Being the innermost and lowest-mass planet, HD\,33142 d shows the largest variation of its eccentricity, which adopts its smallest values when the eccentricities of the other two planets deviate from each other the most (e.g., eccentricity of planet b also small, but that of planet c large). 
Fig.~\ref{fig:ecc2} shows the trajectory evolution of the same stable configuration as a function of the product of their orbital eccentricity evolution and $\sin(\Delta\omega)$. All planet pair geometries exhibit libration around a fixed point, indicative of a mutual apsidal libration.

Given the rather poor constraints of the planet eccentricities and orbital geometries (i.e., their $\omega$) from RV data, we cannot firmly conclude whether the  HD\,33142 system indeed exhibits an apsidal libration, but its overall stability solidifies the existence of the planetary candidates. Yet, apsidally librating configurations are particularly interesting, because such dynamical behavior is likely a result of dynamical evolution, and in the absence of MMRs, are the likely stabilizing mechanism if the true orbits are moderately eccentric. 


\subsection{Fate of the system}
\label{Sec4.5}

HD\,33142 is an early K-giant star, evident to have at least three massive planets in the range of 0.5--2.0 au. 
At the present epoch, we estimate the stellar radius to be $\sim$ 0.02 au, but at the tip of the red giant branch (RGB) the radius is expected to reach up to 1 au. Therefore, the observed planetary system is about to be severely reshaped \citep[see, e.g.,][]{Villaver2009,Veras2016}.

To assess a more realistic long-term dynamical evolution of the HD\,33142 system, we use a custom version of {\sc SyMBA}, which includes second-order dynamical effects such as star-planet tides, stellar evolution, and mass loss. Our {\sc SyMBA} integrator is modified following the prescription of \citet[][and reference therein]{Veras2016}, whereas the stellar evolution input is generated  by the rapid Single-Star Evolution algorithm \citep[SSE,][]{Hurley}. With {\sc SSE} we produce a single evolutionary track starting off the first giant branch track, using as an input our stellar parameter estimates (see \autoref{table_st_param}). At this evolutionary stage, {\sc SSE} started at $\sim$ 2.7\,Gyr (consistent with our estimated age in \autoref{table_st_param}), and evolved the star up to 5\,Gyr, when the star has evolved to a Carbon/Oxygen White Dwarf (WD). Therefore, our N-body test was run for a maximum of 2.2\,Gyr, or until all planets were lost.

\begin{figure}
    \includegraphics[width=\linewidth]{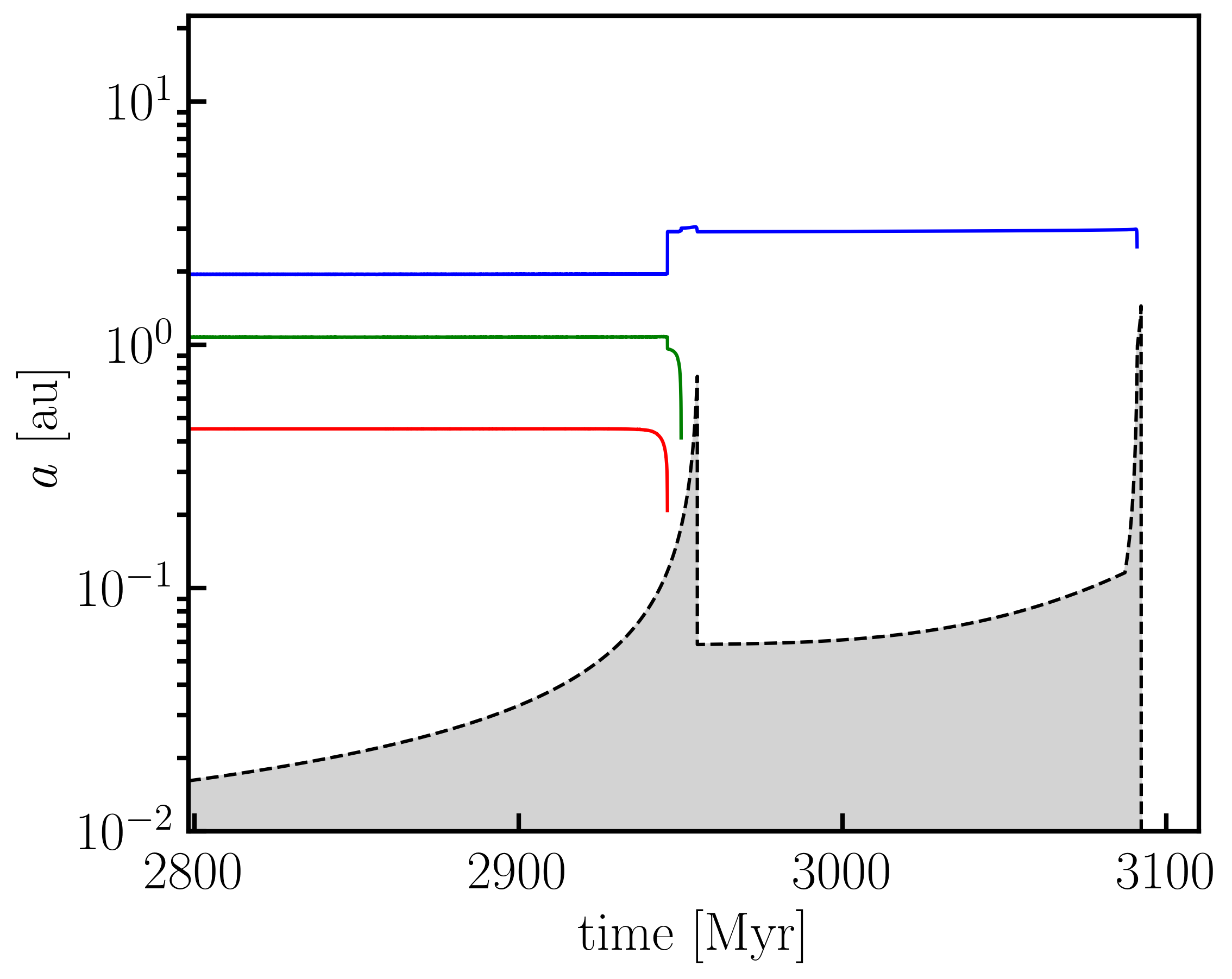} 
 
    \caption{N-body simulations of the HD\,33142 system, including stellar evolution effects on the orbital evolution, in particular the planetary semi-major axes $a$ (colored lines).
    Initial conditions from the N-body simulations were adopted from the stable median posterior values presented in \autoref{table:best_fit}. The evolution of the stellar radius (dashed curve and the underlying gray area) is expressed in astronomical units. The WD phase is off-scale, since the stellar radius at age 3.2 Gyr is only $\sim$4$\times$10$^{-5}$ au.
    }
    \label{Fig:fate}
\end{figure}

Our N-body simulations with stellar evolution show that the HD\,33142 multiple-planet system is only likely to survive less than $\sim$ 150\,Myr from the present epoch. \autoref{Fig:fate} shows the planetary semi-major axes and stellar evolution of the median posterior of stable three-planet configurations. 
The x-axis is the time after the first RGB, whereas the log y-axis shows the semi-major axes and the stellar radius (gray area) expressed in astronomical units. 
Within $\sim$300\,Myr, HD\,33142 ascends through the first RGB, reaches a core helium burning, and “briefly” passes through the Asymptotic Giant Branch (AGB) phases to lose most of its mass to finally settle into a WD remnant.
During the first RGB tip, the stellar radius reaches a maximum of $\sim$0.75\,au, and the star loses $\sim$4\% of its mass. The inflated stellar radius profoundly affects the system stability, thanks to star-planet tidal effects, even before the RGB tip. Our simulations show that the inner two planets HD\,33142\,b \& d will experience a very strong eccentricity and orbital decay before being engulfed before the RGB tip. In contrast, the outermost planet HD\,33142\,c is distant enough to survive the RGB tip phase, after which its orbit will rapidly migrate from $\sim$1.95\,au up to $\sim$2.5\,au, thanks to the stellar mass loss. Yet, HD\,33142\,c will not survive the AGB-tip, during which stellar radius is at its maximum of $\sim$1.45\,au, and tides drag the planet to lower orbits before being engulfed by its dying star.

We test 20 random configurations from the parameter posterior distribution and the stellar mass uncertainty, and we find that the inner two planets are always engulfed at the RGB branch. For the outermost planet HD\,33142\,c, the fate depends on the parameters drawn from the posterior; it is either ejected from the system before the RGB tip, or engulfed in the RGB or the AGB branch. Nonetheless, our numerical simulations with stellar evolution indicate that in all cases the system cannot exist as we observe it once the star reaches the RGB tip.
Detecting planetary companions around HD\,33142 indicate that the star is indeed a first ascent giant star, else the system is very unlikely to exist.\looseness=-4

\section{Discussion and Conclusions}
\label{Sec5}

\begin{figure}[t]
\includegraphics[width=9cm]{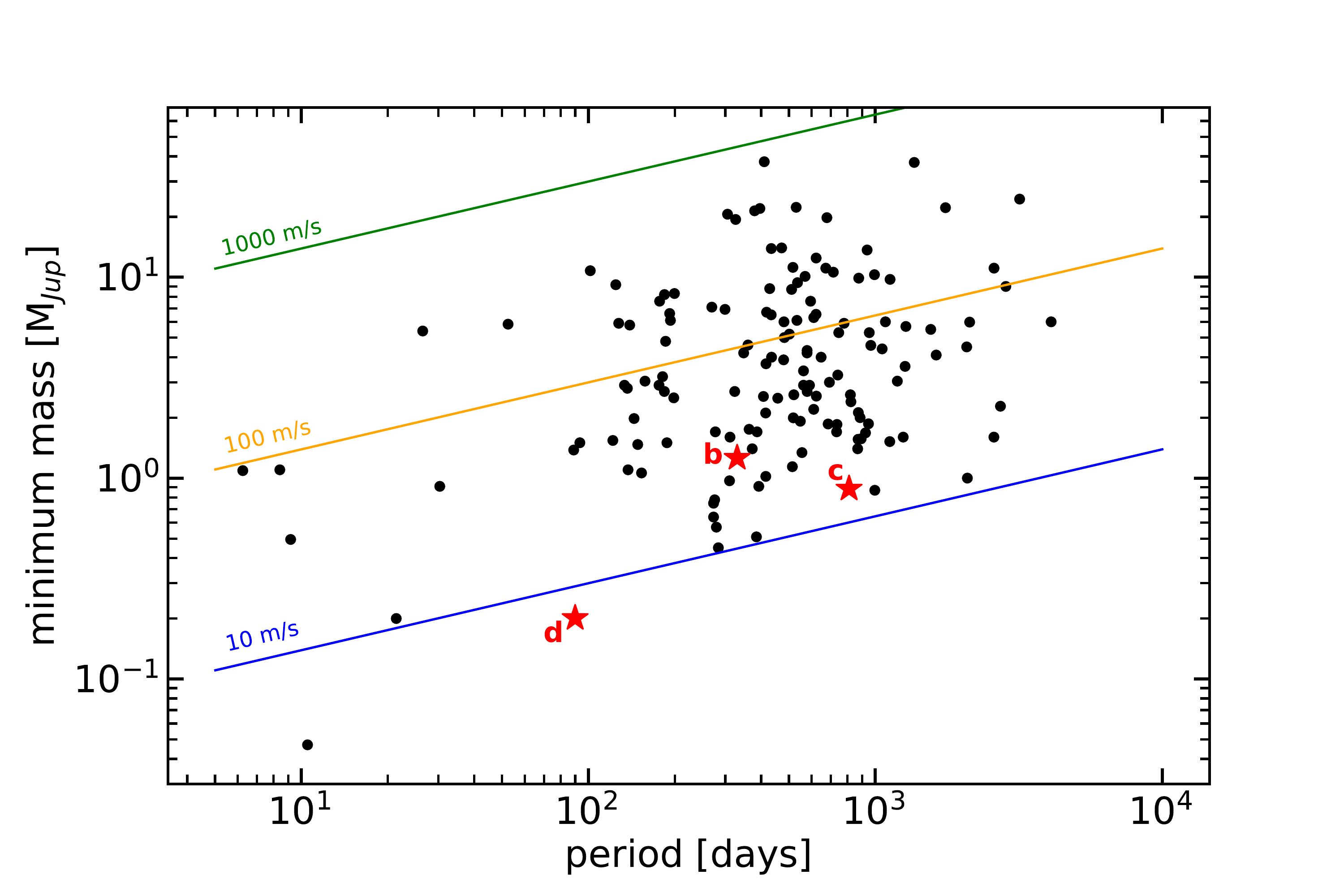}
\caption{Distribution of 141 known exoplanets (black points) around giant stars on a diagram of minimum mass versus orbital period. The planets of the HD~33142 system are indicated as red star symbols. The lines approximate the locations of RV semi-amplitudes of 10, 100 and 1000 m\,s$^{-1}$ for a host star of 1.5~M$_{\odot}$.
}
\label{fig:mass_period_giants}
\end{figure}

\autoref{fig:mass_period_giants} shows the distribution of 141 known exoplanets around giant stars\footnote{Taken from \url{https://www.lsw.uni-heidelberg.de/users/sreffert/giantplanets/giantplanets.php}.} in a diagram of minimum mass versus orbital period. HD\,33142\,b, HD\,33142\,c and HD\,33142\,d are indicated as red stars.  
All three planets are of relatively small mass compared to other planets which have been found around giant stars at comparable periods, indicating that we are pushing detection limits further down. Especially HD\,33142\,c and HD\,33142\,d are the lowest mass-planets known at their respective periods. 
Overall, there are not many planets with periods shorter than about 100 days known around giant stars. There are only two more planets with periods comparable to that of HD\,33142\,d, namely HIP\,67851\,b \citep{Jones2015,Wittenmyer2015,Jones2015b} with a period of 89~days and
8\,UMi\,b \citep{Lee2015} with a period of 93 days, and eight planets with even shorter periods (ranging from 6 to 53 days). These short-period planets around giant stars are so rare that they cannot be efficiently found with radial velocity surveys, but are found in larger numbers only with transit surveys. In fact, the five planets with the smallest periods known to orbit giant stars are all transiting planets.\looseness=-4
 
This is in contrast to planets orbiting main-sequence stars or sub-giant stars, where many hot Jupiters have also been found in radial velocity surveys.
A reason for the dearth of hot Jupiters around giant stars might be stellar evolution and the resulting tidal forces so that close-in planets are eventually engulfed
\citep{Villaver2009}, but it might also have to do with the slightly higher stellar masses in the evolved sample \citep{Currie2009}.\looseness=-4


The formation of the HD\,33142 system is also intriguing.
The HD\,33142  system possesses at least three giant planets, where the innermost planet d is slightly sub-Saturnian in mass, corresponding to a planet-to-star mass ratio of $q=1.41 \times 10^{-4}$. With this mass ratio, planet d would migrate in the type-I regime  \citep[e.g.,][]{Paardekooper2011}, which is mostly directed inwards. However, planet migration can be halted at planet traps, which can either be created through the entropy driven corotation torque \citep[e.g.,][]{Bitsch2013} or through radially increasing gas surface density profiles, which influence the barotropic part of the corotation torque \citep{Masset2006}. At the inner edge of protoplanetary discs, the gas surface density increases initially radially due to a radially decreasing viscosity at the transition between the thermally ionized disc region and the inner edge of the dead zone, giving rise to a planet trap \citep[e.g.,][]{Flock2019}. The thermal ionization depends on the disc's temperature, which in turn depends on the stellar luminosity. For a 1.52 $M_\odot$ star as HD\,33142, this leads to an inner edge at around 0.30-0.35\,au \citep{Flock2016, Flock2019}. However, the planetary trap position is shifted a bit further away from the disc's inner edge, roughly in agreement with the position of planet d in HD\,33142, supporting a smooth migration scenario for the formation of planet d. This is further supported by the relatively low eccentricities of all planets within this system.

Planets b and c orbit the star at around 1 and 2 au, featuring masses of around Jupiter, resulting in star-to-planet mass ratios of around $q\approx 10^{-3}$, indicating a type-II migration \citep{Lin1986} during the gas disc phase. The inner of these two planets, planet b, is also slightly more massive than the outer planet, planet c. Considering that type-II planet migration scales with the planetary mass \citep{Kanagawa2018}, where more massive planets migrate slower, planet c could catch up to planet b, potentially allowing a trapping in resonance. However, this mechanism also depends on the exact disc parameters (e.g., viscosity, H/r, gas surface density), implying that even though the masses of planet b and c are such that the outer planet could catch up with the inner planet, it is not a necessity. 

\citet{Bitsch2020} modeled the formation of giant planet systems from planetary embryos including pebble and gas accretion as well as planetary migration \citep[see also,][]{Matsumura2021}. Their simulations predict that systems harboring gas giants should normally host multiple giant planets, which is also needed to explain the giant planet eccentricity distribution. The low eccentricities for the planets in HD\,33142 indicate that if efficient scattering events between the planets happened, they should have happened during the gas disc phase in order to further damp the eccentricities of the planet to their currently observed values of $\sim$0.1 (\autoref{fig:ecc}). On the other hand, it seems more likely that the system did not undergo scattering events and formed smoothly, where the observed eccentricities are just a result of angular momentum exchange between the planets. The HD\,33142  system featuring three 
giant planets is thus a very interesting example to constrain planet formation.

\section{Summary}
\label{Sec6}

We present an updated orbital solution of the HD\,33142 system based on archival and new observations from FEROS, HARPS, and HIRES. We determined the stellar mass and radius of HD\,33142 to be $M_{\star}$ = 1.52\,$M_{\odot}$ and $R_{\star}$ = 4.17\,$R_{\odot}$, respectively, and found that HD\,33142 is most likely a giant star at the very beginning of the red giant phase, but it could also be at the end of the subgiant branch. We were able to confirm the two already known planets and found a third significant signal at $\sim 90$\,d, which we attribute to a Saturn-mass planet. The  additional new planet is strongly supported by model comparison via Bayesian analysis.  

To confirm that the 90\,d signal is due to a planet, we attempted to compare it to the stellar rotation period $P_\mathrm{rot}$. Our determination of the stellar rotational velocity yielded an upper limit of $v\sin{i} < 2\,{\mathrm{km\,s}}^{-1}$, consistent with the three literature values we found for $P_\mathrm{rot}$. Assuming that the stellar inclination $i=90^\circ $, the $v\sin{i}$ limit of $2\,\mathrm{km\,s}^{-1}$corresponds to a rotational period of $P_\mathrm{rot}=106\,\mathrm{d}$. If the true $v\sin{i}$ is smaller than this limit, then $P_\mathrm{rot}$ is larger. But also, the true value of the inclination $i$ could be smaller so that the true rotational period could also be shorter and encompassing the value of the $90\,\mathrm{d}$ signal we found. This leaves our attempts to determine $P_\mathrm{rot}$ without any conclusive result.

Therefore, we analyzed different activity indicators, none of which shows any significant periodicity near $90\,\mathrm{d}$. This indicates that the most probable explanation for the observed periodicity in the RV time series is the presence of a third planetary companion and not the  rotational modulation of stellar magnetically active regions.\looseness=-4

We determined the orbital parameters of the three planets with a Nelder-Mead (Simplex) algorithm, which optimizes the likelihood of a Keplerian model.  Further, we performed a parameter distribution analysis and estimated the parameter uncertainties by utilizing MCMC and NS sampling methods (see Table \ref{table:best_fit}). The configuration found with the three-planet model consists of two Jupiter-like planets ($m_\mathrm{b}\sin{i} =1.26\,M_{\mathrm{Jup}}$ and $m_\mathrm{c}\sin{i} = 0.89\,M_{\mathrm{Jup}}$) with orbital periods of $P_\mathrm{b} = 330.0\,\mathrm{d}$ and $P_\mathrm{c} = 810.2\,\mathrm{d}$, corresponding to semi-major axes of $a_\mathrm{b} = 1.074\,\mathrm{au}$ and $a_\mathrm{c} = 1.955\,\mathrm{au}$, respectively, as well as an inner Saturn-mass planet ($m_\mathrm{d}\sin{i} = 0.20\,M_{\mathrm{Jup}}$) with a period of $P_\mathrm{d} = 89.9\,\mathrm{d}$ and an orbital separation of $a_\mathrm{d} =  0.452\,\mathrm{au}$.

Contrary to the analysis by \cite{Bryan2016}, our determined mass of the second planet, HD\,33142\,c, is only around one Jupiter-mass, which is one order of magnitude smaller than the originally found and published mass. We assume that this published mass is a  typographical mistake. Also, \cite{Luhn2019} found the smaller planetary mass in their paper, but the signal did not meet their detection criterion, so the higher and wrong mass is still cited in various exoplanet databases.

We analyzed the long-term stability of the HD\,33142 system by performing an N-body integration via the {\sc SyMBA} symplectic integrator, which revealed that the adopted three-planet model, for which a coplanar, edge-on configuration was assumed, is long-term stable. The dynamical properties of the system were performed by selecting 1000 random samples from the NS posteriors and integrating them for 1\,Myr, showing that 66.7\% of the samples were stable over this time base. A smaller set of fits from this sample were tested and confirmed to be long-term stable for at least 100\,Myr. Our dynamical analysis 
showed no evidence of an MMR dynamics between the planet pairs.  Nevertheless, this discovery shows that it is worthwhile to revisit already known planetary systems with sparsely sampled radial velocities, as more precise data over longer temporal baselines lead to more discoveries of individual planets as well as to a more complete picture of the planetary system in which they exist.

\begin{acknowledgments}
This research has made use of the SIMBAD database, operated at CDS, Strasbourg, France.
This work has made use of data from the European Space Agency (ESA) mission
{\it Gaia} (\url{https://www.cosmos.esa.int/gaia}), processed by the {\it Gaia}
Data Processing and Analysis Consortium (DPAC,
\url{https://www.cosmos.esa.int/web/gaia/dpac/consortium}). Funding for the DPAC
has been provided by national institutions, in particular the institutions
participating in the {\it Gaia} Multilateral Agreement.
Funding
for the DPAC has been provided by national institutions, in particular
the institutions participating in the {\it Gaia} Multilateral Agreement.
T.T. acknowledges support by the DFG Research Unit FOR 2544 ”Blue Planets around Red Stars” project No. KU 3625/2-1. 
T.T. further acknowledges support by the BNSF program ``VIHREN-2021'' project No. КП-06-ДВ/5.
S.S.\ and S.R.\ acknowledge support by the DFG Research Unit FOR~2544 {\it Blue  Planets around Red Stars}, 
project no.~RE~2694/4-1. V.W.\ and S.R.\ further acknowledge support by the DFG Priority Program SPP~1992 
{\it Exploring the Diversity of Extrasolar Planets} (RE~2694/5-1).
M.H.L. is supported in part by Hong Kong RGC grant HKU 17305618.
B.B., acknowledges the support of the European Research Council (ERC Starting Grant 757448-PAMDORA).
J.R. acknowledges funding by the DLR (German space agency) via grant 50\,QG\,1403. 
VMP acknowledges financial support from NASA through
grant NNX17AG24G.
V.W.\ and S.R.\ acknowledge the support of the DFG priority program SPP 1992
“Exploring the Diversity of Extrasolar Planets (RE~2694/5-1)”.
We also wish to extend our special thanks to those of Hawaiian ancestry on
whose sacred mountain of Mauna Kea we are privileged to be guests. Without their generous hospitality, the Keck
observations presented herein would not have been possible.

\end{acknowledgments}

\facilities{ESO-3.6m/HARPS, MPG-2.2m/FEROS, KECK-10m/HIRES}

\software{
          Exo-Striker~\citep{Trifonov2019},
          CERES~\citep{Brahm2017},
          emcee~\citep{emcee},
		  dynesty~\citep{Speagle2020},
          spock~\citep{Stock2018}
          }

%

\vspace{5mm}





\clearpage

\bibliography{ref.bib}{}

\begin{thebibliography}{}
\expandafter\ifx\csname natexlab\endcsname\relax\def\natexlab#1{#1}\fi
\providecommand{\url}[1]{\href{#1}{#1}}
\providecommand{\dodoi}[1]{doi:~\href{http://doi.org/#1}{\nolinkurl{#1}}}
\providecommand{\doeprint}[1]{\href{http://ascl.net/#1}{\nolinkurl{http://ascl.net/#1}}}
\providecommand{\doarXiv}[1]{\href{https://arxiv.org/abs/#1}{\nolinkurl{https://arxiv.org/abs/#1}}}

\bibitem[{ESA(1997)}]{ESA1997}
 1997, ESA Special Publication, Vol. 1200, {The HIPPARCOS and TYCHO catalogues.
  Astrometric and photometric star catalogues derived from the ESA HIPPARCOS
  Space Astrometry Mission}

\bibitem[{{Anglada-Escud{\'e}} {et~al.}(2010){Anglada-Escud{\'e}},
  {L{\'o}pez-Morales}, \& {Chambers}}]{Anglada2010}
{Anglada-Escud{\'e}}, G., {L{\'o}pez-Morales}, M., \& {Chambers}, J.~E. 2010,
  The Astrophysical Journal, 709, 168, \dodoi{10.1088/0004-637X/709/1/168}

\bibitem[{Anglada-Escud{\'e} {et~al.}(2016)Anglada-Escud{\'e}, Amado, Barnes,
  Berdi{\~n}as, Butler, Coleman, de~La~Cueva, Dreizler, Endl, Giesers,
  {et~al.}}]{Anglada-Escude2016}
Anglada-Escud{\'e}, G., Amado, P.~J., Barnes, J., {et~al.} 2016, Nature, 536,
  437

\bibitem[{{Arenou} \& {Luri}(1999)}]{Arenou1999}
{Arenou}, F., \& {Luri}, X. 1999, in Astronomical Society of the Pacific
  Conference Series, Vol. 167, Harmonizing Cosmic Distance Scales in a
  Post-HIPPARCOS Era, ed. D.~{Egret} \& A.~{Heck}, 13--32.
\newblock \doarXiv{astro-ph/9812094}

\bibitem[{{Baluev}(2009)}]{Baluev2009}
{Baluev}, R.~V. 2009, Monthly Notices of the Royal Astronomical Society, 393,
  969, \dodoi{10.1111/j.1365-2966.2008.14217.x}

\bibitem[{{Bitsch} {et~al.}(2013){Bitsch}, {Crida}, {Morbidelli}, {Kley}, \&
  {Dobbs-Dixon}}]{Bitsch2013}
{Bitsch}, B., {Crida}, A., {Morbidelli}, A., {Kley}, W., \& {Dobbs-Dixon}, I.
  2013, \aap, 549, A124, \dodoi{10.1051/0004-6361/201220159}

\bibitem[{{Bitsch} {et~al.}(2020){Bitsch}, {Trifonov}, \&
  {Izidoro}}]{Bitsch2020}
{Bitsch}, B., {Trifonov}, T., \& {Izidoro}, A. 2020, \aap, 643, A66,
  \dodoi{10.1051/0004-6361/202038856}

\bibitem[{{Boisvert} {et~al.}(2018){Boisvert}, {Nelson}, \&
  {Steffen}}]{Boisvert2018}
{Boisvert}, J.~H., {Nelson}, B.~E., \& {Steffen}, J.~H. 2018, \mnras, 480,
  2846, \dodoi{10.1093/mnras/sty2023}

\bibitem[{{Brahm} {et~al.}(2017){Brahm}, {Jord{\'a}n}, \&
  {Espinoza}}]{Brahm2017}
{Brahm}, R., {Jord{\'a}n}, A., \& {Espinoza}, N. 2017, Publications of the
  Astronomical Society of the Pacific, 129, 034002,
  \dodoi{10.1088/1538-3873/aa5455}

\bibitem[{{Bressan} {et~al.}(2012){Bressan}, {Marigo}, {Girardi}, {Salasnich},
  {Dal Cero}, {Rubele}, \& {Nanni}}]{Bressan2012}
{Bressan}, A., {Marigo}, P., {Girardi}, L., {et~al.} 2012, \mnras, 427, 127,
  \dodoi{10.1111/j.1365-2966.2012.21948.x}

\bibitem[{{Bryan} {et~al.}(2016){Bryan}, {Knutson}, {Howard}, {Ngo}, {Batygin},
  {Crepp}, {Fulton}, {Hinkley}, {Isaacson}, {Johnson}, {Marcy}, \&
  {Wright}}]{Bryan2016}
{Bryan}, M.~L., {Knutson}, H.~A., {Howard}, A.~W., {et~al.} 2016, The
  Astrophysical Journal, 821, 89, \dodoi{10.3847/0004-637X/821/2/89}

\bibitem[{{Butler} {et~al.}(1996){Butler}, {Marcy}, {Williams}, {McCarthy},
  {Dosanjh}, \& {Vogt}}]{Butler1996}
{Butler}, R.~P., {Marcy}, G.~W., {Williams}, E., {et~al.} 1996, Publications of
  the Astronomical Society of the Pacific, 108, 500, \dodoi{10.1086/133755}

\bibitem[{{Butler} {et~al.}(2017){Butler}, {Vogt}, {Laughlin}, {Burt},
  {Rivera}, {Tuomi}, {Teske}, {Arriagada}, {Diaz}, {Holden}, \&
  {Keiser}}]{Butler2017}
{Butler}, R.~P., {Vogt}, S.~S., {Laughlin}, G., {et~al.} 2017, The Astronomical
  Journal, 153, 208, \dodoi{10.3847/1538-3881/aa66ca}

\bibitem[{{Currie}(2009)}]{Currie2009}
{Currie}, T. 2009, \apjl, 694, L171, \dodoi{10.1088/0004-637X/694/2/L171}

\bibitem[{{Duncan} {et~al.}(1998){Duncan}, {Levison}, \& {Lee}}]{Duncan1998}
{Duncan}, M.~J., {Levison}, H.~F., \& {Lee}, M.~H. 1998, The Astronomical
  Journal, 116, 2067, \dodoi{10.1086/300541}

\bibitem[{{Espinoza} {et~al.}(2019){Espinoza}, {Kossakowski}, \&
  {Brahm}}]{Espinoza2019}
{Espinoza}, N., {Kossakowski}, D., \& {Brahm}, R. 2019, \mnras, 490, 2262,
  \dodoi{10.1093/mnras/stz2688}

\bibitem[{{Flock} {et~al.}(2016){Flock}, {Fromang}, {Turner}, \&
  {Benisty}}]{Flock2016}
{Flock}, M., {Fromang}, S., {Turner}, N.~J., \& {Benisty}, M. 2016, \apj, 827,
  144, \dodoi{10.3847/0004-637X/827/2/144}

\bibitem[{{Flock} {et~al.}(2019){Flock}, {Turner}, {Mulders}, {Hasegawa},
  {Nelson}, \& {Bitsch}}]{Flock2019}
{Flock}, M., {Turner}, N.~J., {Mulders}, G.~D., {et~al.} 2019, \aap, 630, A147,
  \dodoi{10.1051/0004-6361/201935806}

\bibitem[{{Foreman-Mackey} {et~al.}(2013){Foreman-Mackey}, {Hogg}, {Lang}, \&
  {Goodman}}]{emcee}
{Foreman-Mackey}, D., {Hogg}, D.~W., {Lang}, D., \& {Goodman}, J. 2013,
  Publications of the Astronomical Society of the Pacific, 125, 306,
  \dodoi{10.1086/670067}

\bibitem[{{Gaia Collaboration} {et~al.}(2018){Gaia Collaboration}, {Brown},
  {Vallenari}, {Prusti}, {de Bruijne}, {Babusiaux}, {Bailer-Jones}, {Biermann},
  {Evans}, {Eyer}, {Jansen}, {Jordi}, {Klioner}, {Lammers}, {Lindegren},
  {Luri}, {Mignard}, {Panem}, {Pourbaix}, {Randich}, {Sartoretti}, {Siddiqui},
  {Soubiran}, {van Leeuwen}, {Walton}, {Arenou}, {Bastian}, {Cropper},
  {Drimmel}, {Katz}, {Lattanzi}, {Bakker}, {Cacciari}, {Casta{\~n}eda},
  {Chaoul}, {Cheek}, {De Angeli}, {Fabricius}, {Guerra}, {Holl}, {Masana},
  {Messineo}, {Mowlavi}, {Nienartowicz}, {Panuzzo}, {Portell}, {Riello},
  {Seabroke}, {Tanga}, {Th{\'e}venin}, {Gracia-Abril}, {Comoretto},
  {Garcia-Reinaldos}, {Teyssier}, {Altmann}, {Andrae}, {Audard},
  {Bellas-Velidis}, {Benson}, {Berthier}, {Blomme}, {Burgess}, {Busso},
  {Carry}, {Cellino}, {Clementini}, {Clotet}, {Creevey}, {Davidson}, {De
  Ridder}, {Delchambre}, {Dell'Oro}, {Ducourant},
  {Fern{\'a}ndez-Hern{\'a}ndez}, {Fouesneau}, {Fr{\'e}mat}, {Galluccio},
  {Garc{\'\i}a-Torres}, {Gonz{\'a}lez-N{\'u}{\~n}ez}, {Gonz{\'a}lez-Vidal},
  {Gosset}, {Guy}, {Halbwachs}, {Hambly}, {Harrison}, {Hern{\'a}ndez},
  {Hestroffer}, {Hodgkin}, {Hutton}, {Jasniewicz}, {Jean-Antoine-Piccolo},
  {Jordan}, {Korn}, {Krone-Martins}, {Lanzafame}, {Lebzelter}, {L{\"o}ffler},
  {Manteiga}, {Marrese}, {Mart{\'\i}n-Fleitas}, {Moitinho}, {Mora}, {Muinonen},
  {Osinde}, {Pancino}, {Pauwels}, {Petit}, {Recio-Blanco}, {Richards},
  {Rimoldini}, {Robin}, {Sarro}, {Siopis}, {Smith}, {Sozzetti}, {S{\"u}veges},
  {Torra}, {van Reeven}, {Abbas}, {Abreu Aramburu}, {Accart}, {Aerts},
  {Altavilla}, {{\'A}lvarez}, {Alvarez}, {Alves}, {Anderson}, {Andrei},
  {Anglada Varela}, {Antiche}, {Antoja}, {Arcay}, {Astraatmadja}, {Bach},
  {Baker}, {Balaguer-N{\'u}{\~n}ez}, {Balm}, {Barache}, {Barata}, {Barbato},
  {Barblan}, {Barklem}, {Barrado}, {Barros}, {Barstow}, {Bartholom{\'e}
  Mu{\~n}oz}, {Bassilana}, {Becciani}, {Bellazzini}, {Berihuete}, {Bertone},
  {Bianchi}, {Bienaym{\'e}}, {Blanco-Cuaresma}, {Boch}, {Boeche}, {Bombrun},
  {Borrachero}, {Bossini}, {Bouquillon}, {Bourda}, {Bragaglia}, {Bramante},
  {Breddels}, {Bressan}, {Brouillet}, {Br{\"u}semeister}, {Brugaletta},
  {Bucciarelli}, {Burlacu}, {Busonero}, {Butkevich}, {Buzzi}, {Caffau},
  {Cancelliere}, {Cannizzaro}, {Cantat-Gaudin}, {Carballo}, {Carlucci},
  {Carrasco}, {Casamiquela}, {Castellani}, {Castro-Ginard}, {Charlot},
  {Chemin}, {Chiavassa}, {Cocozza}, {Costigan}, {Cowell}, {Crifo}, {Crosta},
  {Crowley}, {Cuypers}, {Dafonte}, {Damerdji}, {Dapergolas}, {David}, {David},
  {de Laverny}, {De Luise}, {De March}, {de Martino}, {de Souza}, {de Torres},
  {Debosscher}, {del Pozo}, {Delbo}, {Delgado}, {Delgado}, {Di Matteo},
  {Diakite}, {Diener}, {Distefano}, {Dolding}, {Drazinos}, {Dur{\'a}n},
  {Edvardsson}, {Enke}, {Eriksson}, {Esquej}, {Eynard Bontemps}, {Fabre},
  {Fabrizio}, {Faigler}, {Falc{\~a}o}, {Farr{\`a}s Casas}, {Federici},
  {Fedorets}, {Fernique}, {Figueras}, {Filippi}, {Findeisen}, {Fonti},
  {Fraile}, {Fraser}, {Fr{\'e}zouls}, {Gai}, {Galleti}, {Garabato},
  {Garc{\'\i}a-Sedano}, {Garofalo}, {Garralda}, {Gavel}, {Gavras}, {Gerssen},
  {Geyer}, {Giacobbe}, {Gilmore}, {Girona}, {Giuffrida}, {Glass}, {Gomes},
  {Granvik}, {Gueguen}, {Guerrier}, {Guiraud}, {Guti{\'e}rrez-S{\'a}nchez},
  {Haigron}, {Hatzidimitriou}, {Hauser}, {Haywood}, {Heiter}, {Helmi}, {Heu},
  {Hilger}, {Hobbs}, {Hofmann}, {Holland}, {Huckle}, {Hypki}, {Icardi},
  {Jan{\ss}en}, {Jevardat de Fombelle}, {Jonker}, {Juh{\'a}sz}, {Julbe},
  {Karampelas}, {Kewley}, {Klar}, {Kochoska}, {Kohley}, {Kolenberg},
  {Kontizas}, {Kontizas}, {Koposov}, {Kordopatis}, {Kostrzewa-Rutkowska},
  {Koubsky}, {Lambert}, {Lanza}, {Lasne}, {Lavigne}, {Le Fustec}, {Le
  Poncin-Lafitte}, {Lebreton}, {Leccia}, {Leclerc}, {Lecoeur-Taibi},
  {Lenhardt}, {Leroux}, {Liao}, {Licata}, {Lindstr{\o}m}, {Lister}, {Livanou},
  {Lobel}, {L{\'o}pez}, {Managau}, {Mann}, {Mantelet}, {Marchal}, {Marchant},
  {Marconi}, {Marinoni}, {Marschalk{\'o}}, {Marshall}, {Martino}, {Marton},
  {Mary}, {Massari}, {Matijevi{\v{c}}}, {Mazeh}, {McMillan}, {Messina},
  {Michalik}, {Millar}, {Molina}, {Molinaro}, {Moln{\'a}r}, {Montegriffo},
  {Mor}, {Morbidelli}, {Morel}, {Morris}, {Mulone}, {Muraveva}, {Musella},
  {Nelemans}, {Nicastro}, {Noval}, {O'Mullane}, {Ord{\'e}novic},
  {Ord{\'o}{\~n}ez-Blanco}, {Osborne}, {Pagani}, {Pagano}, {Pailler},
  {Palacin}, {Palaversa}, {Panahi}, {Pawlak}, {Piersimoni}, {Pineau}, {Plachy},
  {Plum}, {Poggio}, {Poujoulet}, {Pr{\v{s}}a}, {Pulone}, {Racero}, {Ragaini},
  {Rambaux}, {Ramos-Lerate}, {Regibo}, {Reyl{\'e}}, {Riclet}, {Ripepi}, {Riva},
  {Rivard}, {Rixon}, {Roegiers}, {Roelens}, {Romero-G{\'o}mez}, {Rowell},
  {Royer}, {Ruiz-Dern}, {Sadowski}, {Sagrist{\`a} Sell{\'e}s}, {Sahlmann},
  {Salgado}, {Salguero}, {Sanna}, {Santana-Ros}, {Sarasso}, {Savietto},
  {Schultheis}, {Sciacca}, {Segol}, {Segovia}, {S{\'e}gransan}, {Shih},
  {Siltala}, {Silva}, {Smart}, {Smith}, {Solano}, {Solitro}, {Sordo}, {Soria
  Nieto}, {Souchay}, {Spagna}, {Spoto}, {Stampa}, {Steele},
  {Steidelm{\"u}ller}, {Stephenson}, {Stoev}, {Suess}, {Surdej}, {Szabados},
  {Szegedi-Elek}, {Tapiador}, {Taris}, {Tauran}, {Taylor}, {Teixeira},
  {Terrett}, {Teyssand ier}, {Thuillot}, {Titarenko}, {Torra Clotet}, {Turon},
  {Ulla}, {Utrilla}, {Uzzi}, {Vaillant}, {Valentini}, {Valette}, {van Elteren},
  {Van Hemelryck}, {van Leeuwen}, {Vaschetto}, {Vecchiato}, {Veljanoski},
  {Viala}, {Vicente}, {Vogt}, {von Essen}, {Voss}, {Votruba}, {Voutsinas},
  {Walmsley}, {Weiler}, {Wertz}, {Wevers}, {Wyrzykowski}, {Yoldas},
  {{\v{Z}}erjal}, {Ziaeepour}, {Zorec}, {Zschocke}, {Zucker}, {Zurbach}, \&
  {Zwitter}}]{Gaia_Collaboration2018b}
{Gaia Collaboration}, {Brown}, A.~G.~A., {Vallenari}, A., {et~al.} 2018, \aap,
  616, A1, \dodoi{10.1051/0004-6361/201833051}

\bibitem[{{Gaia Collaboration} {et~al.}(2021){Gaia Collaboration}, {Brown},
  {Vallenari}, {Prusti}, {de Bruijne}, {Babusiaux}, {Biermann}, {Creevey},
  {Evans}, {Eyer}, {Hutton}, {Jansen}, {Jordi}, {Klioner}, {Lammers},
  {Lindegren}, {Luri}, {Mignard}, {Panem}, {Pourbaix}, {Randich}, {Sartoretti},
  {Soubiran}, {Walton}, {Arenou}, {Bailer-Jones}, {Bastian}, {Cropper},
  {Drimmel}, {Katz}, {Lattanzi}, {van Leeuwen}, {Bakker}, {Cacciari},
  {Casta{\~n}eda}, {De Angeli}, {Ducourant}, {Fabricius}, {Fouesneau},
  {Fr{\'e}mat}, {Guerra}, {Guerrier}, {Guiraud}, {Jean-Antoine Piccolo},
  {Masana}, {Messineo}, {Mowlavi}, {Nicolas}, {Nienartowicz}, {Pailler},
  {Panuzzo}, {Riclet}, {Roux}, {Seabroke}, {Sordo}, {Tanga}, {Th{\'e}venin},
  {Gracia-Abril}, {Portell}, {Teyssier}, {Altmann}, {Andrae}, {Bellas-Velidis},
  {Benson}, {Berthier}, {Blomme}, {Brugaletta}, {Burgess}, {Busso}, {Carry},
  {Cellino}, {Cheek}, {Clementini}, {Damerdji}, {Davidson}, {Delchambre},
  {Dell'Oro}, {Fern{\'a}ndez-Hern{\'a}ndez}, {Galluccio}, {Garc{\'\i}a-Lario},
  {Garcia-Reinaldos}, {Gonz{\'a}lez-N{\'u}{\~n}ez}, {Gosset}, {Haigron},
  {Halbwachs}, {Hambly}, {Harrison}, {Hatzidimitriou}, {Heiter},
  {Hern{\'a}ndez}, {Hestroffer}, {Hodgkin}, {Holl}, {Jan{\ss}en}, {Jevardat de
  Fombelle}, {Jordan}, {Krone-Martins}, {Lanzafame}, {L{\"o}ffler}, {Lorca},
  {Manteiga}, {Marchal}, {Marrese}, {Moitinho}, {Mora}, {Muinonen}, {Osborne},
  {Pancino}, {Pauwels}, {Petit}, {Recio-Blanco}, {Richards}, {Riello},
  {Rimoldini}, {Robin}, {Roegiers}, {Rybizki}, {Sarro}, {Siopis}, {Smith},
  {Sozzetti}, {Ulla}, {Utrilla}, {van Leeuwen}, {van Reeven}, {Abbas}, {Abreu
  Aramburu}, {Accart}, {Aerts}, {Aguado}, {Ajaj}, {Altavilla}, {{\'A}lvarez},
  {{\'A}lvarez Cid-Fuentes}, {Alves}, {Anderson}, {Anglada Varela}, {Antoja},
  {Audard}, {Baines}, {Baker}, {Balaguer-N{\'u}{\~n}ez}, {Balbinot}, {Balog},
  {Barache}, {Barbato}, {Barros}, {Barstow}, {Bartolom{\'e}}, {Bassilana},
  {Bauchet}, {Baudesson-Stella}, {Becciani}, {Bellazzini}, {Bernet}, {Bertone},
  {Bianchi}, {Blanco-Cuaresma}, {Boch}, {Bombrun}, {Bossini}, {Bouquillon},
  {Bragaglia}, {Bramante}, {Breedt}, {Bressan}, {Brouillet}, {Bucciarelli},
  {Burlacu}, {Busonero}, {Butkevich}, {Buzzi}, {Caffau}, {Cancelliere},
  {C{\'a}novas}, {Cantat-Gaudin}, {Carballo}, {Carlucci}, {Carnerero},
  {Carrasco}, {Casamiquela}, {Castellani}, {Castro-Ginard}, {Castro Sampol},
  {Chaoul}, {Charlot}, {Chemin}, {Chiavassa}, {Cioni}, {Comoretto}, {Cooper},
  {Cornez}, {Cowell}, {Crifo}, {Crosta}, {Crowley}, {Dafonte}, {Dapergolas},
  {David}, {David}, {de Laverny}, {De Luise}, {De March}, {De Ridder}, {de
  Souza}, {de Teodoro}, {de Torres}, {del Peloso}, {del Pozo}, {Delbo},
  {Delgado}, {Delgado}, {Delisle}, {Di Matteo}, {Diakite}, {Diener},
  {Distefano}, {Dolding}, {Eappachen}, {Edvardsson}, {Enke}, {Esquej}, {Fabre},
  {Fabrizio}, {Faigler}, {Fedorets}, {Fernique}, {Fienga}, {Figueras},
  {Fouron}, {Fragkoudi}, {Fraile}, {Franke}, {Gai}, {Garabato},
  {Garcia-Gutierrez}, {Garc{\'\i}a-Torres}, {Garofalo}, {Gavras}, {Gerlach},
  {Geyer}, {Giacobbe}, {Gilmore}, {Girona}, {Giuffrida}, {Gomel}, {Gomez},
  {Gonzalez-Santamaria}, {Gonz{\'a}lez-Vidal}, {Granvik},
  {Guti{\'e}rrez-S{\'a}nchez}, {Guy}, {Hauser}, {Haywood}, {Helmi}, {Hidalgo},
  {Hilger}, {H{\l}adczuk}, {Hobbs}, {Holland}, {Huckle}, {Jasniewicz},
  {Jonker}, {Juaristi Campillo}, {Julbe}, {Karbevska}, {Kervella}, {Khanna},
  {Kochoska}, {Kontizas}, {Kordopatis}, {Korn}, {Kostrzewa-Rutkowska},
  {Kruszy{\'n}ska}, {Lambert}, {Lanza}, {Lasne}, {Le Campion}, {Le Fustec},
  {Lebreton}, {Lebzelter}, {Leccia}, {Leclerc}, {Lecoeur-Taibi}, {Liao},
  {Licata}, {Lindstr{\o}m}, {Lister}, {Livanou}, {Lobel}, {Madrero Pardo},
  {Managau}, {Mann}, {Marchant}, {Marconi}, {Marcos Santos}, {Marinoni},
  {Marocco}, {Marshall}, {Martin Polo}, {Mart{\'\i}n-Fleitas}, {Masip},
  {Massari}, {Mastrobuono-Battisti}, {Mazeh}, {McMillan}, {Messina},
  {Michalik}, {Millar}, {Mints}, {Molina}, {Molinaro}, {Moln{\'a}r},
  {Montegriffo}, {Mor}, {Morbidelli}, {Morel}, {Morris}, {Mulone}, {Munoz},
  {Muraveva}, {Murphy}, {Musella}, {Noval}, {Ord{\'e}novic}, {Orr{\`u}},
  {Osinde}, {Pagani}, {Pagano}, {Palaversa}, {Palicio}, {Panahi}, {Pawlak},
  {Pe{\~n}alosa Esteller}, {Penttil{\"a}}, {Piersimoni}, {Pineau}, {Plachy},
  {Plum}, {Poggio}, {Poretti}, {Poujoulet}, {Pr{\v{s}}a}, {Pulone}, {Racero},
  {Ragaini}, {Rainer}, {Raiteri}, {Rambaux}, {Ramos}, {Ramos-Lerate}, {Re
  Fiorentin}, {Regibo}, {Reyl{\'e}}, {Ripepi}, {Riva}, {Rixon}, {Robichon},
  {Robin}, {Roelens}, {Rohrbasser}, {Romero-G{\'o}mez}, {Rowell}, {Royer},
  {Rybicki}, {Sadowski}, {Sagrist{\`a} Sell{\'e}s}, {Sahlmann}, {Salgado},
  {Salguero}, {Samaras}, {Sanchez Gimenez}, {Sanna}, {Santove{\~n}a},
  {Sarasso}, {Schultheis}, {Sciacca}, {Segol}, {Segovia}, {S{\'e}gransan},
  {Semeux}, {Shahaf}, {Siddiqui}, {Siebert}, {Siltala}, {Slezak}, {Smart},
  {Solano}, {Solitro}, {Souami}, {Souchay}, {Spagna}, {Spoto}, {Steele},
  {Steidelm{\"u}ller}, {Stephenson}, {S{\"u}veges}, {Szabados}, {Szegedi-Elek},
  {Taris}, {Tauran}, {Taylor}, {Teixeira}, {Thuillot}, {Tonello}, {Torra},
  {Torra}, {Turon}, {Unger}, {Vaillant}, {van Dillen}, {Vanel}, {Vecchiato},
  {Viala}, {Vicente}, {Voutsinas}, {Weiler}, {Wevers}, {Wyrzykowski}, {Yoldas},
  {Yvard}, {Zhao}, {Zorec}, {Zucker}, {Zurbach}, \& {Zwitter}}]{Gaia2021}
---. 2021, \aap, 649, A1, \dodoi{10.1051/0004-6361/202039657}

\bibitem[{{Ghezzi} {et~al.}(2018){Ghezzi}, {Montet}, \& {Johnson}}]{Ghezzi2018}
{Ghezzi}, L., {Montet}, B.~T., \& {Johnson}, J.~A. 2018, The Astrophysical
  Journal, 860, 109, \dodoi{10.3847/1538-4357/aac37c}

\bibitem[{{Goodman} \& {Weare}(2010)}]{2010CAMCS...5...65G}
{Goodman}, J., \& {Weare}, J. 2010, Communications in Applied Mathematics and
  Computational Science, 5, 65, \dodoi{10.2140/camcos.2010.5.65}

\bibitem[{{Hara} {et~al.}(2019){Hara}, {Bou{\'e}}, {Laskar}, {Delisle}, \&
  {Unger}}]{Hara2019}
{Hara}, N.~C., {Bou{\'e}}, G., {Laskar}, J., {Delisle}, J.~B., \& {Unger}, N.
  2019, Monthly Notices of the Royal Astronomical Society, 489, 738,
  \dodoi{10.1093/mnras/stz1849}

\bibitem[{{Houk} \& {Smith-Moore}(1988)}]{Houk1988}
{Houk}, N., \& {Smith-Moore}, M. 1988, {Michigan Catalogue of Two-dimensional
  Spectral Types for the HD Stars. Volume 4, Declinations -26.0 to -12.0.},
  Vol.~4

\bibitem[{{Hurley} {et~al.}(2000){Hurley}, {Pols}, \& {Tout}}]{Hurley}
{Hurley}, J.~R., {Pols}, O.~R., \& {Tout}, C.~A. 2000, \mnras, 315, 543,
  \dodoi{10.1046/j.1365-8711.2000.03426.x}

\bibitem[{{Jofr{\'e}} {et~al.}(2015){Jofr{\'e}}, {Petrucci}, {Saffe}, {Saker},
  {Artur de la Villarmois}, {Chavero}, {G{\'o}mez}, \& {Mauas}}]{Jofre2015}
{Jofr{\'e}}, E., {Petrucci}, R., {Saffe}, C., {et~al.} 2015, Astronomy and
  Astrophysics, 574, A50, \dodoi{10.1051/0004-6361/201424474}

\bibitem[{{Johnson} {et~al.}(2011){Johnson}, {Clanton}, {Howard}, {Bowler},
  {Henry}, {Marcy}, {Crepp}, {Endl}, {Cochran}, {MacQueen}, {Wright}, \&
  {Isaacson}}]{Johnson2011}
{Johnson}, J.~A., {Clanton}, C., {Howard}, A.~W., {et~al.} 2011, The
  Astrophysical Journal Supplement, 197, 26, \dodoi{10.1088/0067-0049/197/2/26}

\bibitem[{{Jones} {et~al.}(2015{\natexlab{a}}){Jones}, {Jenkins}, {Rojo},
  {Melo}, \& {Bluhm}}]{Jones2015}
{Jones}, M.~I., {Jenkins}, J.~S., {Rojo}, P., {Melo}, C.~H.~F., \& {Bluhm}, P.
  2015{\natexlab{a}}, Astronomy \& Astrophysics, 573, A3,
  \dodoi{10.1051/0004-6361/201424771}

\bibitem[{{Jones} {et~al.}(2015{\natexlab{b}}){Jones}, {Jenkins}, {Rojo},
  {Olivares}, \& {Melo}}]{Jones2015b}
{Jones}, M.~I., {Jenkins}, J.~S., {Rojo}, P., {Olivares}, F., \& {Melo},
  C.~H.~F. 2015{\natexlab{b}}, \aap, 580, A14,
  \dodoi{10.1051/0004-6361/201525853}

\bibitem[{{Kanagawa} {et~al.}(2018){Kanagawa}, {Tanaka}, \&
  {Szuszkiewicz}}]{Kanagawa2018}
{Kanagawa}, K.~D., {Tanaka}, H., \& {Szuszkiewicz}, E. 2018, \apj, 861, 140,
  \dodoi{10.3847/1538-4357/aac8d9}

\bibitem[{Kass \& Raftery(1995)}]{Kass1995}
Kass, R.~E., \& Raftery, A.~E. 1995, Journal of the american statistical
  association, 90, 773

\bibitem[{{Kaufer} {et~al.}(1999){Kaufer}, {Stahl}, {Tubbesing},
  {N{\o}rregaard}, {Avila}, {Francois}, {Pasquini}, \& {Pizzella}}]{Kaufer1999}
{Kaufer}, A., {Stahl}, O., {Tubbesing}, S., {et~al.} 1999, The Messenger, 95, 8

\bibitem[{{K{\"u}rster} {et~al.}(2015){K{\"u}rster}, {Trifonov}, {Reffert},
  {Kostogryz}, \& {Rodler}}]{Kuerster2015}
{K{\"u}rster}, M., {Trifonov}, T., {Reffert}, S., {Kostogryz}, N.~M., \&
  {Rodler}, F. 2015, Astronomy and Astrophysics, 577, A103,
  \dodoi{10.1051/0004-6361/201525872}

\bibitem[{{K{\"u}rster} {et~al.}(2003){K{\"u}rster}, {Endl}, {Rouesnel}, {Els},
  {Kaufer}, {Brillant}, {Hatzes}, {Saar}, \& {Cochran}}]{Kuerster2003}
{K{\"u}rster}, M., {Endl}, M., {Rouesnel}, F., {et~al.} 2003, Astronomy and
  Astrophysics, 403, 1077, \dodoi{10.1051/0004-6361:20030396}

\bibitem[{{Lee} {et~al.}(2015){Lee}, {Park}, {Lee}, {Jeong}, {Oh}, {Han},
  {Lee}, {Lee}, {Kim}, \& {Kim}}]{Lee2015}
{Lee}, B.~C., {Park}, M.~G., {Lee}, S.~M., {et~al.} 2015, \aap, 584, A79,
  \dodoi{10.1051/0004-6361/201527076}

\bibitem[{{Lee} \& {Peale}(2003)}]{Lee2003}
{Lee}, M.~H., \& {Peale}, S.~J. 2003, The Astrophysical Journal, 592, 1201,
  \dodoi{10.1086/375857}

\bibitem[{{Lin} \& {Papaloizou}(1986)}]{Lin1986}
{Lin}, D.~N.~C., \& {Papaloizou}, J. 1986, \apj, 309, 846,
  \dodoi{10.1086/164653}

\bibitem[{{Luhn} {et~al.}(2019){Luhn}, {Bastien}, {Wright}, {Johnson},
  {Howard}, \& {Isaacson}}]{Luhn2019}
{Luhn}, J.~K., {Bastien}, F.~A., {Wright}, J.~T., {et~al.} 2019, The
  Astronomical Journal, 157, 149, \dodoi{10.3847/1538-3881/aaf5d0}

\bibitem[{{Luque} {et~al.}(2019){Luque}, {Pall{\'e}}, {Kossakowski},
  {Dreizler}, {Kemmer}, {Espinoza}, {Burt}, {Anglada-Escud{\'e}}, {B{\'e}jar},
  {Caballero}, {Collins}, {Collins}, {Cort{\'e}s-Contreras},
  {D{\'\i}ez-Alonso}, {Feng}, {Hatzes}, {Hellier}, {Henning}, {Jeffers},
  {Kaltenegger}, {K{\"u}rster}, {Madden}, {Molaverdikhani}, {Montes}, {Narita},
  {Nowak}, {Ofir}, {Oshagh}, {Parviainen}, {Quirrenbach}, {Reffert}, {Reiners},
  {Rodr{\'\i}guez-L{\'o}pez}, {Schlecker}, {Stock}, {Trifonov}, {Winn},
  {Zapatero Osorio}, {Zechmeister}, {Amado}, {Anderson}, {Batalha}, {Bauer},
  {Bluhm}, {Burke}, {Butler}, {Caldwell}, {Chen}, {Crane}, {Dragomir},
  {Dressing}, {Dynes}, {Jenkins}, {Kaminski}, {Klahr}, {Kotani}, {Lafarga},
  {Latham}, {Lewin}, {McDermott}, {Monta{\~n}{\'e}s-Rodr{\'\i}guez}, {Morales},
  {Murgas}, {Nagel}, {Pedraz}, {Ribas}, {Ricker}, {Rowden}, {Seager},
  {Shectman}, {Tamura}, {Teske}, {Twicken}, {Vanderspeck}, {Wang}, \&
  {Wohler}}]{Luque2019}
{Luque}, R., {Pall{\'e}}, E., {Kossakowski}, D., {et~al.} 2019, \aap, 628, A39,
  \dodoi{10.1051/0004-6361/201935801}

\bibitem[{{Malla} {et~al.}(2020){Malla}, {Stello}, {Huber}, {Montet},
  {Bedding}, {Fredslund Andersen}, {Grundahl}, {Jessen-Hansen}, {Hey}, {Palle},
  {Deng}, {Zhang}, {Chen}, {Lloyd}, \& {Antoci}}]{Malla2020}
{Malla}, S.~P., {Stello}, D., {Huber}, D., {et~al.} 2020, \mnras, 496, 5423,
  \dodoi{10.1093/mnras/staa1793}

\bibitem[{{Mardling}(2013)}]{Mardling2013}
{Mardling}, R.~A. 2013, Monthly Notices of the Royal Astronomical Society, 435,
  2187, \dodoi{10.1093/mnras/stt1438}

\bibitem[{{Masset} {et~al.}(2006){Masset}, {Morbidelli}, {Crida}, \&
  {Ferreira}}]{Masset2006}
{Masset}, F.~S., {Morbidelli}, A., {Crida}, A., \& {Ferreira}, J. 2006, \apj,
  642, 478, \dodoi{10.1086/500967}

\bibitem[{{Matsumura} {et~al.}(2021){Matsumura}, {Brasser}, \&
  {Ida}}]{Matsumura2021}
{Matsumura}, S., {Brasser}, R., \& {Ida}, S. 2021, \aap, 650, A116,
  \dodoi{10.1051/0004-6361/202039210}

\bibitem[{{Mayor} \& {Queloz}(1995)}]{Mayor1995}
{Mayor}, M., \& {Queloz}, D. 1995, Nature, 378, 355, \dodoi{10.1038/378355a0}

\bibitem[{{Mayor} {et~al.}(2003){Mayor}, {Pepe}, {Queloz}, {Bouchy},
  {Rupprecht}, {Lo Curto}, {Avila}, {Benz}, {Bertaux}, {Bonfils}, {Dall},
  {Dekker}, {Delabre}, {Eckert}, {Fleury}, {Gilliotte}, {Gojak}, {Guzman},
  {Kohler}, {Lizon}, {Longinotti}, {Lovis}, {Megevand}, {Pasquini}, {Reyes},
  {Sivan}, {Sosnowska}, {Soto}, {Udry}, {van Kesteren}, {Weber}, \&
  {Weilenmann}}]{Mayor2003}
{Mayor}, M., {Pepe}, F., {Queloz}, D., {et~al.} 2003, The Messenger, 114, 20

\bibitem[{Nelder \& Mead(1965)}]{NelderMead1965}
Nelder, J.~A., \& Mead, R. 1965, Computer Journal, 7, 308

\bibitem[{{Paardekooper} {et~al.}(2011){Paardekooper}, {Baruteau}, \&
  {Kley}}]{Paardekooper2011}
{Paardekooper}, S.~J., {Baruteau}, C., \& {Kley}, W. 2011, \mnras, 410, 293,
  \dodoi{10.1111/j.1365-2966.2010.17442.x}

\bibitem[{{Pepe} {et~al.}(2002){Pepe}, {Mayor}, {Galland}, {Naef}, {Queloz},
  {Santos}, {Udry}, \& {Burnet}}]{2002A&A...388..632P}
{Pepe}, F., {Mayor}, M., {Galland}, F., {et~al.} 2002, \aap, 388, 632,
  \dodoi{10.1051/0004-6361:20020433}

\bibitem[{{Perdelwitz} {et~al.}(2021){Perdelwitz}, {Mittag}, {Tal-Or},
  {Schmitt}, {Caballero}, {Jeffers}, {Reiners}, {Schweitzer}, {Trifonov},
  {Ribas}, {Quirrenbach}, {Amado}, {Seifert}, {Cifuentes},
  {Cort{\'e}s-Contreras}, {Montes}, {Revilla}, \&
  {Skrzypinski}}]{Perdelwitz2021}
{Perdelwitz}, V., {Mittag}, M., {Tal-Or}, L., {et~al.} 2021, \aap, 652, A116,
  \dodoi{10.1051/0004-6361/202140889}

\bibitem[{{Queloz}(1995)}]{1995IAUS..167..221Q}
{Queloz}, D. 1995, in New Developments in Array Technology and Applications,
  ed. A.~G.~D. {Philip}, K.~{Janes}, \& A.~R. {Upgren}, Vol. 167, 221

\bibitem[{{Reiners} {et~al.}(2012){Reiners}, {Joshi}, \&
  {Goldman}}]{2012AJ....143...93R}
{Reiners}, A., {Joshi}, N., \& {Goldman}, B. 2012, \aj, 143, 93,
  \dodoi{10.1088/0004-6256/143/4/93}

\bibitem[{{Ricker} {et~al.}(2015){Ricker}, {Winn}, {Vanderspek}, {Latham},
  {Bakos}, {Bean}, {Berta-Thompson}, {Brown}, {Buchhave}, {Butler}, {Butler},
  {Chaplin}, {Charbonneau}, {Christensen-Dalsgaard}, {Clampin}, {Deming},
  {Doty}, {De Lee}, {Dressing}, {Dunham}, {Endl}, {Fressin}, {Ge}, {Henning},
  {Holman}, {Howard}, {Ida}, {Jenkins}, {Jernigan}, {Johnson}, {Kaltenegger},
  {Kawai}, {Kjeldsen}, {Laughlin}, {Levine}, {Lin}, {Lissauer}, {MacQueen},
  {Marcy}, {McCullough}, {Morton}, {Narita}, {Paegert}, {Palle}, {Pepe},
  {Pepper}, {Quirrenbach}, {Rinehart}, {Sasselov}, {Sato}, {Seager},
  {Sozzetti}, {Stassun}, {Sullivan}, {Szentgyorgyi}, {Torres}, {Udry}, \&
  {Villasenor}}]{Ricker2015}
{Ricker}, G.~R., {Winn}, J.~N., {Vanderspek}, R., {et~al.} 2015, Journal of
  Astronomical Telescopes, Instruments, and Systems, 1, 014003,
  \dodoi{10.1117/1.JATIS.1.1.014003}

\bibitem[{{Skilling}(2004)}]{Skilling2004}
{Skilling}, J. 2004, in American Institute of Physics Conference Series, Vol.
  735, Bayesian Inference and Maximum Entropy Methods in Science and
  Engineering: 24th International Workshop on Bayesian Inference and Maximum
  Entropy Methods in Science and Engineering, ed. R.~{Fischer}, R.~{Preuss}, \&
  U.~V. {Toussaint}, 395--405, \dodoi{10.1063/1.1835238}

\bibitem[{Sneden(1973)}]{s1973}
Sneden, C. 1973, PhD thesis, Univ. Texas, Austin

\bibitem[{{Speagle}(2020)}]{Speagle2020}
{Speagle}, J.~S. 2020, Monthly Notices of the Royal Astronomical Society, 493,
  3132, \dodoi{10.1093/mnras/staa278}

\bibitem[{{Stock} {et~al.}(2018){Stock}, {Reffert}, \&
  {Quirrenbach}}]{Stock2018}
{Stock}, S., {Reffert}, S., \& {Quirrenbach}, A. 2018, \aap, 616, A33,
  \dodoi{10.1051/0004-6361/201833111}

\bibitem[{{Stock} {et~al.}(2020){Stock}, {Kemmer}, {Reffert}, {Trifonov},
  {Kaminski}, {Dreizler}, {Quirrenbach}, {Caballero}, {Reiners}, {Jeffers},
  {Anglada-Escud{\'e}}, {Ribas}, {Amado}, {Barrado}, {Barnes}, {Bauer},
  {Berdi{\~n}as}, {B{\'e}jar}, {Coleman}, {Cort{\'e}s-Contreras},
  {D{\'\i}ez-Alonso}, {Dom{\'\i}nguez-Fern{\'a}ndez}, {Espinoza}, {Haswell},
  {Hatzes}, {Henning}, {Jenkins}, {Jones}, {Kossakowski}, {K{\"u}rster},
  {Lafarga}, {Lee}, {L{\'o}pez Gonz{\'a}lez}, {Montes}, {Morales}, {Morales},
  {Pall{\'e}}, {Pedraz}, {Rodr{\'\i} guez}, {Rodr{\'\i}guez-L{\'o}pez}, \&
  {Zechmeister}}]{Stock2020}
{Stock}, S., {Kemmer}, J., {Reffert}, S., {et~al.} 2020, arXiv e-prints,
  arXiv:2002.01772.
\newblock \doarXiv{2002.01772}

\bibitem[{{Tody}(1986)}]{Tody1986}
{Tody}, D. 1986, in Society of Photo-Optical Instrumentation Engineers (SPIE)
  Conference Series, Vol. 627, Instrumentation in astronomy VI, ed. D.~L.
  {Crawford}, 733, \dodoi{10.1117/12.968154}

\bibitem[{{Trifonov}(2019)}]{ES2019}
{Trifonov}, T. 2019, {The Exo-Striker: Transit and radial velocity interactive
  fitting tool for orbital analysis and N-body simulations}.
\newblock \doeprint{1906.004}

\bibitem[{{Trifonov} {et~al.}(2019){Trifonov}, {Stock}, {Henning}, {Reffert},
  {K{\"u}rster}, {Lee}, {Bitsch}, {Butler}, \& {Vogt}}]{Trifonov2019}
{Trifonov}, T., {Stock}, S., {Henning}, T., {et~al.} 2019, The Astronomical
  Journal, 157, 93, \dodoi{10.3847/1538-3881/aafa11}

\bibitem[{{Trifonov} {et~al.}(2020){Trifonov}, {Lee}, {K{\"u}rster}, {Henning},
  {Grishin}, {Stock}, {Tjoa}, {Caballero}, {Wong}, {Bauer}, {Quirrenbach},
  {Zechmeister}, {Ribas}, {Reffert}, {Reiners}, {Amado}, {Kossakowski},
  {Azzaro}, {B{\'e}jar}, {Cort{\'e}s-Contreras}, {Dreizler}, {Hatzes},
  {Jeffers}, {Kaminski}, {Lafarga}, {Montes}, {Morales}, {Pavlov},
  {Rodr{\'\i}guez-L{\'o}pez}, {Schmitt}, {Solano}, \& {Barnes}}]{Trifonov2020}
{Trifonov}, T., {Lee}, M.~H., {K{\"u}rster}, M., {et~al.} 2020, Astronomy \&
  Astrophysics, 638, A16, \dodoi{10.1051/0004-6361/201936987}

\bibitem[{Trifonov {et~al.}(2021)Trifonov, {Caballero}, {Morales}, {Seifahrt},
  {Ribas}, {Reiners}, {Bean}, {Luque}, {Parviainen}, {Pall{\'e}}, {Stock},
  {Zechmeister}, {Amado}, {Anglada-Escud{\'e}}, {Azzaro}, {Barclay},
  {B{\'e}jar}, {Bluhm}, {Casasayas-Barris}, {Cifuentes}, {Collins}, {Collins},
  {Cort{\'e}s-Contreras}, {de Leon}, {Dreizler}, {Dressing}, {Esparza-Borges},
  {Espinoza}, {Fausnaugh}, {Fukui}, {Hatzes}, {Hellier}, {Henning}, {Henze},
  {Herrero}, {Jeffers}, {Jenkins}, {Jensen}, {Kaminski}, {Kasper},
  {Kossakowski}, {K{\"u}rster}, {Lafarga}, {Latham}, {Mann}, {Molaverdikhani},
  {Montes}, {Montet}, {Murgas}, {Narita}, {Oshagh}, {Passegger}, {Pollacco},
  {Quinn}, {Quirrenbach}, {Ricker}, {Rodr{\'\i}guez L{\'o}pez}, {Sanz-Forcada},
  {Schwarz}, {Schweitzer}, {Seager}, {Shporer}, {Stangret}, {St{\"u}rmer},
  {Tan}, {Tenenbaum}, {Twicken}, {Vanderspek}, \& {Winn}}]{Trifonov2021a}
Trifonov, T., {Caballero}, J.~A., {Morales}, J.~C., {et~al.} 2021, Science,
  371, 1038, \dodoi{10.1126/science.abd7645}

\bibitem[{{Trifonov} {et~al.}(2021){Trifonov}, {Brahm}, {Espinoza}, {Henning},
  {Jord{\'a}n}, {Nesvorny}, {Dawson}, {Lissauer}, {Lee}, {Kossakowski},
  {Rojas}, {Hobson}, {Sarkis}, {Schlecker}, {Bitsch}, {Bakos}, {Barbieri},
  {Bhatti}, {Butler}, {Crane}, {Nandakumar}, {D{\'\i}az}, {Shectman}, {Teske},
  {Torres}, {Suc}, {Vines}, {Wang}, {Ricker}, {Shporer}, {Vanderburg},
  {Dragomir}, {Vanderspek}, {Burke}, {Daylan}, {Shiao}, {Jenkins}, {Wohler},
  {Seager}, \& {Winn}}]{Trifonov2021b}
{Trifonov}, T., {Brahm}, R., {Espinoza}, N., {et~al.} 2021, \aj, 162, 283,
  \dodoi{10.3847/1538-3881/ac1bbe}

\bibitem[{{Trotta}(2008)}]{Trotta2008}
{Trotta}, R. 2008, Contemporary Physics, 49, 71,
  \dodoi{10.1080/00107510802066753}

\bibitem[{{Valenti} \& {Piskunov}(1996)}]{Valenti1996}
{Valenti}, J.~A., \& {Piskunov}, N. 1996, Astronomy and Astrophysics
  Supplement, 118, 595

\bibitem[{{van Leeuwen}(2007)}]{VanLeeuwen2007}
{van Leeuwen}, F. 2007, \aap, 474, 653, \dodoi{10.1051/0004-6361:20078357}

\bibitem[{{Veras}(2016)}]{Veras2016}
{Veras}, D. 2016, \mnras, 463, 2958, \dodoi{10.1093/mnras/stw2170}

\bibitem[{{Villaver} \& {Livio}(2009)}]{Villaver2009}
{Villaver}, E., \& {Livio}, M. 2009, \apjl, 705, L81,
  \dodoi{10.1088/0004-637X/705/1/L81}

\bibitem[{{Vogt} {et~al.}(1994){Vogt}, {Allen}, {Bigelow}, {Bresee}, {Brown},
  {Cantrall}, {Conrad}, {Couture}, {Delaney}, {Epps}, {Hilyard}, {Hilyard},
  {Horn}, {Jern}, {Kanto}, {Keane}, {Kibrick}, {Lewis}, {Osborne},
  {Pardeilhan}, {Pfister}, {Ricketts}, {Robinson}, {Stover}, {Tucker}, {Ward},
  \& {Wei}}]{Vogt1994}
{Vogt}, S.~S., {Allen}, S.~L., {Bigelow}, B.~C., {et~al.} 1994, in Society of
  Photo-Optical Instrumentation Engineers (SPIE) Conference Series, Vol. 2198,
  Instrumentation in Astronomy VIII, ed. D.~L. {Crawford} \& E.~R. {Craine},
  362, \dodoi{10.1117/12.176725}

\bibitem[{{Wisdom} \& {Holman}(1991)}]{Wisdom1991}
{Wisdom}, J., \& {Holman}, M. 1991, Astronomical Journal, 102, 1528,
  \dodoi{10.1086/115978}

\bibitem[{{Wittenmyer} {et~al.}(2015){Wittenmyer}, {Wang}, {Liu}, {Horner},
  {Endl}, {Johnson}, {Tinney}, \& {Carter}}]{Wittenmyer2015}
{Wittenmyer}, R.~A., {Wang}, L., {Liu}, F., {et~al.} 2015, The Astrophysical
  Journal, 800, 74, \dodoi{10.1088/0004-637X/800/1/74}

\bibitem[{{Wittenmyer} {et~al.}(2013){Wittenmyer}, {Wang}, {Horner}, {Tinney},
  {Butler}, {Jones}, {O'Toole}, {Bailey}, {Carter}, {Salter}, {Wright}, \&
  {Zhou}}]{Wittenmyer2013}
{Wittenmyer}, R.~A., {Wang}, S., {Horner}, J., {et~al.} 2013, The Astrophysical
  Journal Supplement, 208, 2, \dodoi{10.1088/0067-0049/208/1/2}

\bibitem[{{Zechmeister} \& {K{\"u}rster}(2009)}]{Zechmeister2009}
{Zechmeister}, M., \& {K{\"u}rster}, M. 2009, Astronomy and Astrophysics, 496,
  577, \dodoi{10.1051/0004-6361:200811296}

\bibitem[{{Zechmeister} {et~al.}(2018){Zechmeister}, {Reiners}, {Amado},
  {Azzaro}, {Bauer}, {B{\'e}jar}, {Caballero}, {Guenther}, {Hagen}, {Jeffers},
  {Kaminski}, {K{\"u}rster}, {Launhardt}, {Montes}, {Morales}, {Quirrenbach},
  {Reffert}, {Ribas}, {Seifert}, {Tal-Or}, \& {Wolthoff}}]{Zechmeister2018}
{Zechmeister}, M., {Reiners}, A., {Amado}, P.~J., {et~al.} 2018, Astronomy and
  Astrophysics, 609, A12, \dodoi{10.1051/0004-6361/201731483}

\bibitem[{{Zechmeister} {et~al.}(2019){Zechmeister}, {Dreizler}, {Ribas},
  {Reiners}, {Caballero}, {Bauer}, {B{\'e}jar}, {Gonz{\'a}lez-Cuesta},
  {Herrero}, {Lalitha}, {L{\'o}pez-Gonz{\'a}lez}, {Luque}, {Morales},
  {Pall{\'e}}, {Rodr{\'\i}guez}, {Rodr{\'\i}guez L{\'o}pez}, {Tal-Or},
  {Anglada-Escud{\'e}}, {Quirrenbach}, {Amado}, {Abril}, {Aceituno},
  {Aceituno}, {Alonso-Floriano}, {Ammler-von Eiff}, {Antona Jim{\'e}nez},
  {Anwand-Heerwart}, {Arroyo-Torres}, {Azzaro}, {Baroch}, {Barrado},
  {Becerril}, {Ben{\'\i}tez}, {Berdi{\~n}as}, {Bergond}, {Bluhm},
  {Brinkm{\"o}ller}, {del Burgo}, {Calvo Ortega}, {Cano}, {Cardona
  Guill{\'e}n}, {Carro}, {C{\'a}rdenas V{\'a}zquez}, {Casal},
  {Casasayas-Barris}, {Casanova}, {Chaturvedi}, {Cifuentes}, {Claret},
  {Colom{\'e}}, {Cort{\'e}s-Contreras}, {Czesla}, {D{\'\i}ez-Alonso}, {Dorda},
  {Fern{\'a}ndez}, {Fern{\'a}ndez-Mart{\'\i}n}, {Fuhrmeister}, {Fukui},
  {Galad{\'\i}-Enr{\'\i}quez}, {Gallardo Cava}, {Garcia de la Fuente},
  {Garcia-Piquer}, {Garc{\'\i}a Vargas}, {Gesa}, {G{\'o}ngora Rueda},
  {Gonz{\'a}lez-{\'A}lvarez}, {Gonz{\'a}lez Hern{\'a}ndez},
  {Gonz{\'a}lez-Peinado}, {Gr{\"o}zinger}, {Gu{\`a}rdia}, {Guijarro}, {de
  Guindos}, {Hatzes}, {Hauschildt}, {Hedrosa}, {Helmling}, {Henning},
  {Hermelo}, {Hern{\'a}ndez Arabi}, {Hern{\'a}ndez Casta{\~n}o}, {Hern{\'a}ndez
  Otero}, {Hintz}, {Huke}, {Huber}, {Jeffers}, {Johnson}, {de Juan},
  {Kaminski}, {Kemmer}, {Kim}, {Klahr}, {Klein}, {Kl{\"u}ter}, {Klutsch},
  {Kossakowski}, {K{\"u}rster}, {Labarga}, {Lafarga}, {Llamas}, {Lamp{\'o}n},
  {Lara}, {Launhardt}, {L{\'a}zaro}, {Lodieu}, {L{\'o}pez del Fresno},
  {L{\'o}pez-Puertas}, {L{\'o}pez Salas}, {L{\'o}pez-Santiago}, {Mag{\'a}n
  Madinabeitia}, {Mall}, {Mancini}, {Mand el}, {Marfil}, {Mar{\'\i}n Molina},
  {Maroto Fern{\'a}ndez}, {Mart{\'\i}n}, {Mart{\'\i}n-Fern{\'a}ndez},
  {Mart{\'\i}n-Ruiz}, {Marvin}, {Mirabet}, {Monta{\~n}{\'e}s-Rodr{\'\i}guez},
  {Montes}, {Moreno-Raya}, {Nagel}, {Naranjo}, {Narita}, {Nortmann}, {Nowak},
  {Ofir}, {Oshagh}, {Panduro}, {Parviainen}, {Pascual}, {Passegger}, {Pavlov},
  {Pedraz}, {P{\'e}rez-Calpena}, {P{\'e}rez Medialdea}, {Perger}, {Perryman},
  {Rabaza}, {Ram{\'o}n Ballesta}, {Rebolo}, {Redondo}, {Reffert}, {Reinhardt},
  {Rhode}, {Rix}, {Rodler}, {Rodr{\'\i}guez Trinidad}, {Rosich}, {Sadegi},
  {S{\'a}nchez-Blanco}, {S{\'a}nchez Carrasco}, {S{\'a}nchez-L{\'o}pez},
  {Sanz-Forcada}, {Sarkis}, {Sarmiento}, {Sch{\"a}fer}, {Schmitt},
  {Sch{\"o}fer}, {Schweitzer}, {Seifert}, {Shulyak}, {Solano}, {Sota}, {Stahl},
  {Stock}, {Strachan}, {Stuber}, {St{\"u}rmer}, {Su{\'a}rez}, {Tabernero},
  {Tala Pinto}, {Trifonov}, {Veredas}, {Vico Linares}, {Vilardell}, {Wagner},
  {Wolthoff}, {Xu}, {Yan}, \& {Zapatero Osorio}}]{Zechmeister2019}
{Zechmeister}, M., {Dreizler}, S., {Ribas}, I., {et~al.} 2019, \aap, 627, A49,
  \dodoi{10.1051/0004-6361/201935460}

\end{thebibliography}
\bibliographystyle{aasjournal}

\begin{appendix} 

\label{appendix}

In this Appendix, in \autoref{fig:CP} we show the three-planet posterior distribution achieved from our NS sampling test of the combined FEROS, HARPS, and HIRES RV data of HD\,33142.
 \autoref{tab:FEROS_RVdata} show the FEROS radial velocity and activity index measurements. 
  \autoref{tab:HARPS_RVdata} show the HARPS radial velocity and activity index measurements for HD\,33142, derived with the DRS and \texttt{serval} pipelines.
  \autoref{tab:HIRES_RVdata} show the HIRES radial velocity and activity index measurements.

 \setcounter{table}{0}
\renewcommand{\thetable}{A\arabic{table}}

\setcounter{figure}{0}
\renewcommand{\thefigure}{A\arabic{figure}}

\begin{figure*}
    \centering
    \includegraphics[width=\textwidth]{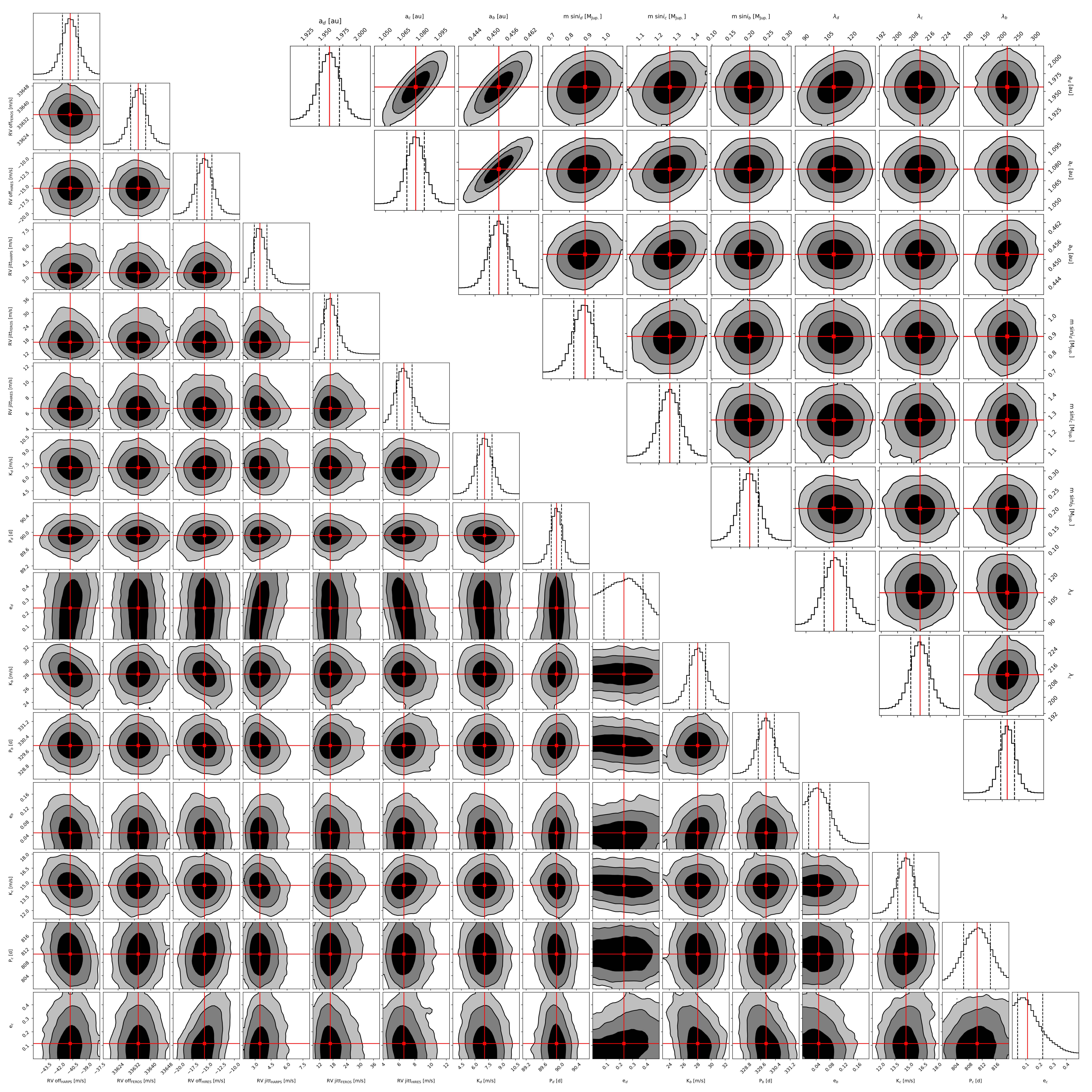}
    \caption{Corner plots showing the posteriors of all parameters from the NS analysis for the three-planet model. The top panel of every column shows the histogram distribution of each parameter. The bottom left cornerplot, from top to bottom and left to right the parameters are: RV offsets of HARPS, FEROS, and HIRES, RV jitters of HARPS, FEROS, and HIRES, $K_\mathrm{d}$, $P_\mathrm{d}$, $e_\mathrm{d}$,  $K_\mathrm{b}$, $P_\mathrm{b}$, $e_\mathrm{b}$,  $K_\mathrm{c}$, $P_\mathrm{c}$, $e_\mathrm{c}$. The top right cornerplot shows the posterior distribution of the derived parameters; the semi-major axes $a_\mathrm{d}$, $a_\mathrm{b}$, $a_\mathrm{c}$, the minimum masses $m_\mathrm{d} \sin i$, $m_\mathrm{b} \sin i$, $m_\mathrm{c} \sin i$,  and the mean longitudes, which are $\lambda_\mathrm{b,c,d}$ = $\omega_\mathrm{b,c,d}$ + $M_\mathrm{db,c,d}$. The two-dimensional contours indicate the 1-, 2- and 3-$\sigma$ confidence levels of the posterior distribution. The red lines mark the position of the median values.}
    \label{fig:CP}
\end{figure*}

\newpage

\begin{table}
\caption{Radial velocity and activity indices of HD\,33142 obtained from FEROS.}
\label{tab:FEROS_RVdata}

\centering
\resizebox{\textwidth}{!}{
\begin{tabular}{@{}c c c c c c c c c c c c c@{}}

\hline\hline
\noalign{\vskip 0.5mm}

Epoch [BJD] & RV [m\,s$^{-1}$] & $\sigma_{RV}$ [m\,s$^{-1}$]  &  H$\alpha$ & $\sigma_{H\alpha}$  &  NaI D1 & $\sigma_{\mathrm NaI D1}$  &  NaI D2 & $\sigma_{\mathrm NaI D2}$ &  CaII H  & $\sigma_{\mathrm CaII H}$ & CaII K  & $\sigma_{\mathrm CaII K}$   \\
\hline

 2455879.83295   &  33738.80   &   3.10   & 0.8334  & 0.0006   & 0.7359  & 0.0009  &  0.8127  & 0.0010  &  0.1200  & 0.0018  & 0.1668  & 0.0248 \\
 2457956.93368   &  33590.50   &   4.20   & 0.8407  & 0.0009   & 0.7351  & 0.0015  &  0.8046  & 0.0016  &  0.1150  & 0.0054  & 0.1081  & 0.0088 \\
 2457967.90668   &  33618.00   &   5.50   & 0.8830  & 0.0013   & 0.7437  & 0.0023  &  0.8094  & 0.0023  &  0.1105  & 0.0088  & 0.1265  & 0.0131 \\
 2457973.91266   &  33610.90   &   6.80   & 0.8698  & 0.0018   & 0.7444  & 0.0032  &  0.8108  & 0.0032  &  0.1008  & 0.0163  & 0.1724  & 0.0238 \\
 2457979.90407   &  33582.60   &   4.10   & 0.8433  & 0.0008   & 0.7366  & 0.0014  &  0.8105  & 0.0015  &  0.1178  & 0.0038  & 0.1499  & 0.0067 \\
 2457981.92253   &  33594.40   &   4.10   & 0.8400  & 0.0008   & 0.7376  & 0.0014  &  0.8103  & 0.0015  &  0.1103  & 0.0034  & 0.1467  & 0.0064 \\
 2457983.89550   &  33599.50   &   4.00   & 0.8409  & 0.0008   & 0.7368  & 0.0014  &  0.8061  & 0.0014  &  0.1167  & 0.0034  & 0.1549  & 0.0062 \\
 2457985.90556   &  33601.50   &   4.80   & 0.8522  & 0.0010   & 0.7423  & 0.0018  &  0.8088  & 0.0019  &  0.1037  & 0.0054  & 0.1146  & 0.0084 \\
 2458019.89399   &  37211.70   &   4.10   & \dots   & \dots    & \dots   & \dots   & \dots    & \dots   & \dots    & \dots   & \dots   & \dots  \\
 2458036.72940   &  33643.00   &   4.50   & 0.8469  & 0.0010   & 0.7379  & 0.0017  &  0.8108  & 0.0017  &  0.1061  & 0.0056  & 0.1175  & 0.0094 \\
 2458054.76927   &  33631.20   &   5.10   & 0.8529  & 0.0012   & 0.7404  & 0.0021  &  0.8103  & 0.0021  &  0.1071  & 0.0059  & 0.1430  & 0.0090 \\
 2458061.66286   &  33638.20   &   4.10   & 0.8552  & 0.0009   & 0.7341  & 0.0014  &  0.8048  & 0.0015  &  0.0886  & 0.0049  & 0.0882  & 0.0086 \\
 2458094.79396   &  33964.80   &  40.50   & \dots   & \dots    & \dots   & \dots   & \dots    & \dots   & \dots    & \dots   & \dots   & \dots  \\
 2458100.69229   &  33645.60   &   4.40   & 0.8504  & 0.0010   & 0.7406  & 0.0016  &  0.8106  & 0.0017  &  0.1138  & 0.0039  & 0.1408  & 0.0074 \\
 2458127.64249   &  33700.00   &   4.10   & 0.8592  & 0.0009   & 0.7558  & 0.0014  &  0.7978  & 0.0014  & \dots   &  \dots   & \dots   & \dots  \\
 2458127.64511   &  33720.10   &   4.10   & 0.8380  & 0.0009   & 0.7428  & 0.0014  &  0.8017  & 0.0014  & \dots    & \dots   & \dots   & \dots  \\
 2458134.70082   &  33671.00   &   4.30   & 0.8484  & 0.0009   & 0.7395  & 0.0015  &  0.8112  & 0.0015  &  0.0996  & 0.0048  & 0.0769  & 0.0084 \\
 2458153.59075   &  33676.00   &   4.00   & 0.8531  & 0.0008   & 0.7204  & 0.0013  &  0.7972  & 0.0013  &  0.1038  & 0.0029  & 0.1241  & 0.0062 \\
 2458199.53727   &  33669.80   &   4.30   & 0.8592  & 0.0010   & 0.7353  & 0.0016  &  0.8134  & 0.0016  &  0.0894  & 0.0055  & 0.0727  & 0.0095 \\
 2458199.53925   &  33670.80   &   4.30   & 0.8545  & 0.0009   & 0.7351  & 0.0015  &  0.8094  & 0.0015  &  0.0653  & 0.0056  & 0.0639  & 0.0110 \\
 2458199.54123   &  33677.70   &   4.30   & 0.8557  & 0.0009   & 0.7370  & 0.0015  &  0.8099  & 0.0016  &  0.0745  & 0.0052  & 0.0696  & 0.0090 \\
 2458568.90575   &  33606.30   &   3.70   & 0.8402  & 0.0007   & 0.7349  & 0.0012  &  0.8071  & 0.0012  &  0.1197  & 0.0041  & 0.1334  & 0.0085 \\
 2458684.90575   &  33613.80   &   4.00   & 0.8352  & 0.0008   & 0.7362  & 0.0014  &  0.8117  & 0.0014  &  0.1311  & 0.0141  & 0.0653  & 0.0235 \\
 2458732.91062   &  33639.90   &  12.10   & 0.8565  & 0.0042   & 0.7467  & 0.0080  &  0.8027  & 0.0079  &  0.0395  & 0.0500  & 0.0006  & 0.0819 \\
 2458761.87144   &  33627.30   &   3.90   & 0.8350  & 0.0008   & 0.7417  & 0.0014  &  0.8106  & 0.0014  &  0.1329  & 0.0027  & 0.1579  & 0.0053 \\
 2458787.76438   &  33648.20   &   4.20   & 0.8345  & 0.0009   & 0.7385  & 0.0015  &  0.8101  & 0.0015  &  0.1269  & 0.0036  & 0.1740  & 0.0067 \\
 2458814.74357   &  33628.00   &   3.90   & 0.8354  & 0.0008   & 0.7386  & 0.0013  &  0.8101  & 0.0013  &  0.1276  & 0.0026  & 0.1802  & 0.0055 \\

\hline
\noalign{\vskip 0.5mm}

\hline

\end{tabular}}

\end{table}

\begin{table}
\caption{Radial velocity and activity indices of HD\,33142 obtained from HARPS.}
\label{tab:HARPS_RVdata}

\centering
\resizebox{\textwidth}{!}{
\begin{tabular}{c c c c c c c c c c c c c c c c c c c c c}

\hline\hline
\noalign{\vskip 0.5mm}

Epoch [BJD] & RV [m\,s$^{-1}$] & $\sigma_{RV}$ [m\,s$^{-1}$]  & H$\alpha$ & $\sigma_{H\alpha}$  &  NaD$_1$ & $\sigma_{\rm NaD_1}$  &  NaD$_2$ & $\sigma_{\rm NaD_2}$ & CRX [m\,s$^{-1}$]&   $\sigma_{\rm CRX} [{\rm m\,s^{-1}}]$  & dLW [m$^{2}$\,s$^{-2}$] & $\sigma_{\rm dLW} [m^{2}\,s^{-2}]$  &   BIS [km\,s$^{-1}$]  &  $\sigma_{\rm BIS}$ & FWHM [km\,s$^{-1}$] & $\sigma_{\rm FWHM}$   &  Cont & $\sigma_{\rm Cont}$  \\

\hline     
\noalign{\vskip 0.5mm} 
2457609.91389&-55.087&0.536&0.46007&0.00065&0.24569&0.00072&0.34572&0.00096&9.18&8.34&7.73&1.0&0.0472& 0.001& 6.7041& 0.001& 46.612&0.01\\ 
2457609.92015&-50.815&0.61&0.46156&0.00074&0.24667&0.00082&0.34644&0.00109&9.96&8.31&10.18&1.24&0.0474& 0.001& 6.7061&0.001&46.594&0.01\\
2457609.92659&-51.801&0.524&0.46037&0.00064&0.24598&0.00071&0.34483&0.00093&2.17&7.29&8.65&1.11&0.0451& 0.001&6.7044&0.001&46.607&0.01\\
2457857.50461&-22.713&0.671&0.45733&0.00073&0.24532&0.00084&0.34598&0.00112&2.64&9.1&0.03&1.25&0.0458& 0.001& 6.7016&0.001&46.63&0.01\\
2457857.51052&-19.795&0.72&0.45746&0.00078&0.2481&0.00091&0.34245&0.0012&-0.38&10.02&-0.39&1.33&0.0458& 0.001& 6.7018&0.001&46.643&0.01\\
2457857.51678&-18.097&0.781&0.4586&0.00084&0.24619&0.00098&0.34683&0.0013&17.9&9.84&1.44&1.63&0.0474& 0.001& 6.704&0.001&46.619&0.01\\
2458003.89463&-70.376&0.609&0.4593&0.00073&0.24584&0.0008&0.33804&0.00106&-7.45&8.01&7.1&1.3&0.0467& 0.001& 6.6995&0.001&46.604&0.01\\
2458003.90066&-69.609&0.582&0.45975&0.00069&0.24798&0.00076&0.3378&0.00101&-16.45&8.52&9.67&1.12&0.0469& 0.001& 6.7018&0.001&46.598&0.01\\
2458003.90693&-70.762&0.535&0.45961&0.00063&0.24699&0.00069&0.33903&0.00092&-8.45&7.88&7.72&1.03&0.0448& 0.001& 6.7039&0.001&46.591&0.01\\
2458084.7618&-37.507&0.483&0.45699&0.00053&0.24706&0.0006&0.34536&0.00079&6.41&7.01&3.85&1.03&0.0487& 0.001& 6.7027&0.001&46.63&0.01\\
2458084.76624&-40.069&0.471&0.45669&0.00052&0.2404&0.00058&0.34587&0.00077&0.07&5.95&1.36&0.91&0.0517& 0.001& 6.7023&0.001&46.638&0.01\\
2458084.77076&-39.975&0.466&0.45827&0.00051&0.24013&0.00057&0.34661&0.00076&4.56&7.13&-0.15&0.84&0.0487& 0.001& 6.6991&0.001&46.652&0.01\\
2458143.52478&-4.854&0.462&0.46291&0.00052&0.2477&0.0006&0.34303&0.00079&-9.99&6.05&1.36&0.83&0.0471& 0.001& 6.6985&0.001&46.636&0.01\\
2458143.53099&-3.796&0.449&0.46342&0.00051&0.24781&0.00058&0.34328&0.00076&-10.49&5.31&-3.08&0.79&0.0486& 0.001& 6.6979&0.001&46.661&0.01\\
2458143.53698&-4.34&0.44&0.46204&0.0005&0.24844&0.00057&0.33461&0.00075&-10.17&5.21&-3.03&0.86&0.0522& 0.001& 6.6998&0.001&46.654&0.01\\
2458144.60014&-6.945&0.625&0.45672&0.00072&0.24044&0.00082&0.34377&0.00109&8.22&6.65&-1.64&1.36&0.0499& 0.001& 6.6986&0.001&46.66&0.01\\
2458144.60629&-6.087&0.604&0.45855&0.00069&0.24051&0.00079&0.34458&0.00105&20.03&6.33&-4.19&1.14&0.047& 0.001& 6.7019&0.001&46.668&0.01\\
2458144.61246&-2.652&0.589&0.45767&0.00067&0.24113&0.00077&0.34303&0.00101&14.27&7.96&-4.51&1.2&0.0477& 0.001& 6.6957&0.001&46.675&0.01\\
2458145.62584&-11.183&0.513&0.4614&0.00052&0.24809&0.00062&0.33149&0.00081&-7.83&5.72&-1.09&1.04&0.0494& 0.001& 6.6969&0.001&46.664&0.01\\
2458145.632&-8.017&0.596&0.46247&0.00061&0.24741&0.00073&0.33228&0.00095&0.48&7.72&-4.66&1.3&0.0497& 0.001& 6.6957&0.001&46.642&0.01\\
2458145.63804&-1.541&0.529&0.46213&0.00054&0.24767&0.00064&0.3344&0.00084&6.01&5.21&-5.43&1.05&0.0487& 0.001& 6.6977&0.001&46.665&0.01\\
2458172.5288&-4.578&0.869&0.46326&0.00103&0.24284&0.00122&0.34194&0.00159&-8.01&6.31&-5.32&1.82&0.0494& 0.001& 6.6991&0.001&46.66&0.01\\
2458173.52798&-4.177&0.472&0.46255&0.00053&0.24204&0.00061&0.34118&0.00081&19.1&9.55&-1.63&1.01&0.0503& 0.001& 6.6998&0.001&46.644&0.01\\
2458174.5198&-7.179&0.456&0.45753&0.0005&0.24046&0.00059&0.34136&0.00077&-2.49&5.62&-2.41&0.93&0.0496& 0.001& 6.6961&0.001&46.657&0.01\\
2458175.52104&-4.748&0.492&0.4568&0.00055&0.24745&0.00064&0.34105&0.00084&-8.96&5.37&0.74&0.96&0.0485& 0.001& 6.6981&0.001&46.627&0.01\\
2458191.52248&1.418&0.523&0.45693&0.00058&0.24757&0.00068&0.33324&0.0009&-4.4&6.46&-4.12&1.04&0.0492& 0.001& 6.7007&0.001&46.654&0.01\\
2458191.52857&2.233&0.465&0.45774&0.0005&0.24888&0.00059&0.33277&0.00078&-1.28&5.89&0.09&0.99&0.0493& 0.001&6.7019&0.001&46.635&0.01\\
2458191.53454&2.841&0.455&0.45721&0.00049&0.2481&0.00058&0.33175&0.00076&-5.04&5.58&0.98&1.02&0.0474& 0.001&6.7001&0.001&46.625&0.01\\
2458194.50983&0.889&0.347&0.46136&0.00038&0.24244&0.00045&0.34218&0.00059&3.53&5.18&-0.65&0.79&0.0495& 0.001&6.6989&0.001&46.639&0.01\\
2458196.53961&-1.871&0.479&0.46133&0.0005&0.2471&0.0006&0.34223&0.00079&5.25&5.14&1.88&0.88&0.0479& 0.001&6.7003&0.001&46.636&0.01\\
2458196.54581&-0.8&0.472&0.46161&0.00048&0.24733&0.00058&0.34268&0.00076&3.62&6.13&1.72&0.85&0.0489& 0.001& 6.7014&0.001&46.631&0.01\\
2458196.55196&-0.398&0.478&0.46158&0.00047&0.24805&0.00058&0.34484&0.00076&6.87&5.63&0.45&0.84&0.05& 0.001& 6.7019&0.001&46.621&0.01\\
2458221.48152&1.844&0.56&0.45804&0.00058&0.24138&0.00071&0.34225&0.00093&-0.38&6.35&-0.38&1.1&0.0489& 0.001& 6.7016&0.001&46.631&0.01\\
2458221.48809&2.316&0.781&0.45814&0.0008&0.24039&0.00099&0.34074&0.00129&-1.38&7.44&-4.19&1.31&0.0506& 0.001& 6.7003&0.001&46.654&0.01\\
2458221.49424&-1.599&0.596&0.45792&0.00063&0.24305&0.00077&0.34118&0.001&10.44&6.54&-1.2&1.1&0.0495& 0.001& 6.7013&0.001&46.646&0.01\\
2458222.47187&1.498&0.469&0.45782&0.00049&0.24076&0.00058&0.34407&0.00077&-0.4&6.14&2.96&0.95&0.0499& 0.001& 6.7042&0.001&46.625&0.01\\
2458222.47808&-2.313&0.46&0.45679&0.00047&0.24038&0.00057&0.34426&0.00075&2.35&6.3&5.2&0.96&0.0506& 0.001& 6.7065&0.001&46.601&0.01\\
2458222.48424&-8.119&0.485&0.45766&0.00049&0.24099&0.0006&0.3441&0.00079&-0.91&6.15&5.2&1.03&0.0502& 0.001& 6.7059&0.001&46.596&0.01\\
2458224.51695&-4.11&0.479&0.4621&0.00048&0.23996&0.00058&0.34471&0.00077&1.8&5.72&2.68&0.95&0.0518& 0.001& 6.7044&0.001&46.629&0.01\\
2458224.52305&-5.048&0.484&0.46197&0.00048&0.2411&0.00059&0.34531&0.00078&3.55&5.3&2.53&0.94&0.0466& 0.001& 6.7067&0.001&46.625&0.01\\
2458224.52909&-7.203&0.505&0.4613&0.0005&0.24022&0.00061&0.34487&0.00081&-0.45&5.39&-0.28&0.98&0.0483& 0.001& 6.7044&0.001&6.638&0.01\\
2458225.48043&-13.121&0.691&0.45711&0.00073&0.24107&0.00089&0.3437&0.00117&0.38&6.94&-1.46&1.38&0.0499& 0.001& 6.7006&0.001&46.634&0.01\\
2458225.48634&-12.128&0.664&0.45802&0.00075&0.24052&0.0009&0.34274&0.00119&-20.52&6.67&-0.71&1.14&0.0548& 0.001& 6.703&0.001&46.642&0.01\\
2458225.49256&-9.144&0.588&0.45585&0.00064&0.2395&0.00077&0.34327&0.00102&-15.62&7.34&0.1&1.12&0.0501& 0.001& 6.7044&0.001&46.65&0.01\\
2458318.91454&-64.604&0.923&0.46026&0.00102&0.2409&0.00123&0.3487&0.00163&6.12&9.54&2.38&1.84&0.0449& 0.001& 6.6978&0.001&46.615&0.01\\
2458318.92302&-62.893&1.03&0.4594&0.00114&0.2384&0.00137&0.3487&0.00182&7.63&11.01&-3.85&2.66&0.05& 0.001& 6.6919&0.001&46.717&0.01\\
2458318.93149&-54.605&0.918&0.45955&0.00103&0.23983&0.00123&0.33755&0.00163&5.16&9.6&2.0&1.93&0.0443& 0.001& 6.7012&0.001&46.637&0.01\\
2458383.81627&-47.063&0.573&0.46077&0.00071&0.2464&0.00082&0.33796&0.00108&9.65&6.44&1.63&1.11&0.0521& 0.001& 6.6986&0.001&46.651&0.01\\
2458383.82225&-47.035&0.639&0.45913&0.00077&0.24645&0.00089&0.34025&0.00119&1.13&7.58&0.23&1.28&0.051& 0.001& 6.6969&0.001&46.659&0.01\\
2458383.82853&-42.457&0.596&0.46054&0.00075&0.24741&0.00086&0.33758&0.00113&-12.23&6.27&-1.51&1.16&0.0486& 0.001& 6.7002&0.001&46.655&0.01\\
2458385.78696&-48.793&0.511&0.4607&0.00062&0.24605&0.00071&0.34548&0.00094&0.69&5.01&1.61&0.91&0.0483& 0.001& 6.6967&0.001&46.656&0.01\\
2458385.79305&-49.663&0.511&0.45977&0.00063&0.2454&0.00072&0.34632&0.00096&-0.37&5.86&3.61&1.14&0.0462& 0.001& 6.6992&0.001&46.645&0.01\\
2458385.79925&-48.085&0.503&0.4607&0.00062&0.24629&0.00071&0.34497&0.00094&1.28&6.31&2.44&1.03&0.048& 0.001& 6.6965&0.001&46.638&0.01\\
2458403.86168&-39.066&1.238&0.45839&0.00145&0.24152&0.00175&0.34842&0.0023&12.92&13.05&-6.37&2.83&0.0508& 0.001& 6.7033&0.001&46.638&0.01\\
2458403.86779&-36.524&1.219&0.45769&0.00143&0.24096&0.00172&0.34843&0.00227&19.34&12.47&-1.68&2.8&0.0436& 0.001& 6.6968&0.001&46.676&0.01\\
2458403.87389&-36.053&0.983&0.45671&0.00116&0.2378&0.00137&0.34879&0.00182&7.09&11.59&-4.3&2.03&0.0485& 0.001& 6.6989&0.001&46.704&0.01\\
2458416.76954&-40.104&0.518&0.46295&0.00063&0.23618&0.00072&0.33518&0.00096&4.38&7.98&4.92&1.15&0.0478& 0.001& 6.7003&0.001&46.615&0.01\\
2458416.7758&-46.856&0.475&0.46287&0.00058&0.23467&0.00066&0.33377&0.00088&-9.47&8.96&3.89&0.98&0.051& 0.001& 6.701&0.001&46.625&0.01\\
2458416.78188&-47.025&0.472&0.45813&0.00057&0.2353&0.00065&0.33376&0.00087&3.72&7.2&0.15&0.94&0.0527& 0.001& 6.6993&0.001&46.657&0.01\\
2458416.78802&-39.86&0.483&0.45812&0.00059&0.23612&0.00067&0.3421&0.00089&-7.32&8.41&-1.42&1.0&0.0491& 0.001& 6.6963&0.001&46.647&0.01\\
2458436.85486&-38.765&0.597&0.45588&0.00068&0.24767&0.00081&0.34701&0.00107&3.28&9.52&-0.29&1.12&0.05& 0.001& 6.7005&0.001&46.642&0.01\\
2458436.86084&-43.375&0.585&0.4565&0.00067&0.24426&0.00078&0.34567&0.00104&2.06&8.5&2.66&1.27&0.0485& 0.001& 6.7007&0.001&46.654&0.01\\
2458451.75009&-34.039&0.507&0.46009&0.00058&0.2471&0.00068&0.34306&0.0009&9.85&8.66&-3.46&1.02&0.0466& 0.001& 6.6977&0.001&46.679&0.01\\
2458451.75394&-33.031&0.507&0.46082&0.00058&0.24866&0.00068&0.3411&0.00089&6.56&7.93&-3.88&0.78&0.0506& 0.001& 6.6949&0.001&46.672&0.01\\
2458451.75778&-32.882&0.514&0.45977&0.00058&0.24759&0.00069&0.3441&0.0009&-6.82&8.64&-2.32&1.08&0.0508& 0.001& 6.6976&0.001&46.657&0.01\\
2458454.68064&-26.925&1.109&0.45662&0.0014&0.23751&0.00163&0.34588&0.00216&2.74&11.9&-2.41&2.49&0.0465& 0.001& 6.7002&0.001&46.667&0.01\\
2458454.68554&-26.785&0.825&0.45555&0.00104&0.24281&0.00121&0.34203&0.00158&8.85&8.46&-0.03&1.76&0.0466& 0.001& 6.6993&0.001&46.653&0.01\\
2458454.69161&-27.52&0.908&0.45626&0.00115&0.24049&0.00134&0.34572&0.00177&0.03&10.4&-1.7&1.92&0.0523& 0.001& 6.7036&0.001&46.645&0.01\\
2458474.68296&-22.581&0.428&0.4569&0.00048&0.24747&0.00057&0.34502&0.00075&5.2&7.68&0.75&0.84&0.0512& 0.001& 6.6967&0.001&46.643&0.01\\
2458474.68911&-19.549&0.435&0.46155&0.00049&0.24848&0.00058&0.34241&0.00076&0.84&7.42&0.15&1.01&0.0493& 0.001& 6.6992&0.001&46.635&0.01\\
2458474.69527&-16.181&0.414&0.46149&0.00047&0.24833&0.00055&0.34179&0.00072&3.31&6.7&-0.73&0.85&0.0517& 0.001& 6.6949&0.001&46.644&0.01\\
2458476.78879&-13.542&0.539&0.45943&0.00057&0.24278&0.00069&0.33484&0.00091&6.99&6.02&-0.08&0.95&0.0503& 0.001& 6.6997&0.001&46.646&0.01\\
2458476.79495&-14.734&0.59&0.46026&0.00062&0.24167&0.00075&0.34226&0.00099&4.79&6.38&0.55&1.17&0.0483& 0.001& 6.7046&0.001&46.628&0.01\\
2458476.80116&-15.852&0.598&0.46122&0.00061&0.24193&0.00075&0.34149&0.00099&24.35&6.53&2.56&1.14&0.0524& 0.001& 6.7021&0.001&46.619&0.01\\
2458529.65501&-43.613&0.613&0.45715&0.0007&0.24347&0.00082&0.34119&0.00108&-4.02&6.34&-7.81&1.31&0.0475& 0.001& 6.6973&0.001&46.652&0.01\\
2458529.6603&-44.267&1.034&0.45644&0.00107&0.24143&0.00132&0.33918&0.00173&4.41&10.92&-9.29&2.55&0.0493& 0.001& 6.6985&0.001&46.596&0.01\\
2458529.66725&-42.333&0.727&0.45785&0.00083&0.24161&0.00097&0.34284&0.00128&3.22&6.54&-8.63&1.51&0.0461& 0.001& 6.7019&0.001&46.631&0.01\\
2458531.63395&-39.172&0.481&0.4564&0.00056&0.24825&0.00065&0.34096&0.00085&-3.68&6.43&-0.07&1.2&0.0487& 0.001& 6.7007&0.001&46.631&0.01\\
2458531.64044&-40.523&0.483&0.45709&0.00056&0.24958&0.00065&0.33839&0.00085&-10.22&6.38&0.63&0.98&0.049& 0.001& 6.7023&0.001&46.63&0.01\\
2458531.64635&-41.45&0.511&0.4563&0.00059&0.24837&0.00069&0.33736&0.0009&-13.89&5.65&-1.57&1.1&0.0453& 0.001& 6.6977&0.001&6.65&0.01\\
2458538.59208&-39.652&0.457&0.45647&0.00051&0.24384&0.00059&0.33589&0.00076&-2.43&7.27&-3.43&1.12&0.05& 0.001& 6.699&0.001&46.648&0.01\\
2458538.59813&-38.996&0.465&0.4565&0.00051&0.24385&0.0006&0.33731&0.00077&-3.43&6.21&-1.17&0.94&0.0478& 0.001& 6.7004&0.001&46.638&0.01\\
2458538.60521&-41.482&0.581&0.45669&0.00066&0.24918&0.00076&0.32887&0.00099&-11.19&7.99&-6.49&1.37&0.0517& 0.001& 6.6988&0.001&46.661&0.01\\
2458542.55889&-31.866&0.551&0.45696&0.00067&0.248&0.00076&0.33804&0.00099&-9.13&7.23&-3.93&0.98&0.0491& 0.001& 6.6971&0.001&46.641&0.01\\
2458542.56498&-38.659&0.511&0.45636&0.00062&0.24881&0.0007&0.33895&0.00091&5.03&7.08&-3.19&1.03&0.0505& 0.001& 6.6971&0.001&46.653&0.01\\
2458542.57131&-43.288&0.524&0.45676&0.00063&0.24947&0.00071&0.33646&0.00093&4.32&7.16&-5.69&0.99&0.0456& 0.001& 6.6966&0.001&46.654&0.01\\
2458546.57545&-42.877&0.542&0.45739&0.00064&0.24902&0.00073&0.33884&0.00095&0.73&8.2&-2.03&1.05&0.0506& 0.001& 6.7002&0.001&46.641&0.01\\
2458546.58156&-41.172&0.556&0.45716&0.00067&0.24976&0.00075&0.33816&0.00098&-1.92&8.63&-2.38&1.19&0.0509& 0.001& 6.6988&0.001&46.661&0.01\\
2458546.58783&-36.919&0.562&0.45671&0.00067&0.2491&0.00076&0.33899&0.00099&-7.32&8.59&-2.69&1.14&0.0469& 0.001& 6.6991&0.001&46.643&0.01\\
2458550.59229&-35.537&0.487&0.45829&0.00054&0.24973&0.00062&0.33619&0.0008&-3.23&5.61&0.27&0.92&0.0456& 0.001& 6.701&0.001&46.628&0.01\\
2458550.59857&-33.797&0.516&0.45849&0.00057&0.2509&0.00066&0.3364&0.00085&-6.5&5.02&2.05&1.02&0.0468& 0.001& 6.7016&0.001&46.615&0.01\\
2458550.60438&-35.497&0.576&0.45783&0.00064&0.24918&0.00074&0.33647&0.00095&-5.1&7.3&0.23&1.4&0.0445& 0.001& 6.699&0.001&46.644&0.01\\
2458565.49203&-36.998&0.493&0.45894&0.00056&0.24171&0.00062&0.33903&0.00081&-15.62&6.65&0.07&0.96&0.0475& 0.001& 6.7008&0.001&46.617&0.01\\
2458565.49707&-34.357&0.508&0.45759&0.00058&0.24124&0.00064&0.34074&0.00084&-5.71&6.13&-2.04&0.89&0.0475& 0.001& 6.7018&0.001&46.644&0.01\\
2458565.50203&-35.379&0.498&0.45704&0.00056&0.24275&0.00063&0.3408&0.00082&-4.27&7.68&1.61&0.92&0.0485& 0.001& 6.7024&0.001&46.633&0.01\\
2458754.89592&-39.67&0.837&0.46179&0.00107&0.25238&0.00114&0.34839&0.00149&4.82&8.6&2.6&1.76&0.0493& 0.001& 6.7006& 0.001&46.612&0.01\\
2458754.90211&-40.056&0.844&0.46178&0.00107&0.25233&0.00114&0.34988&0.0015&1.32&8.32&1.59&1.76&0.0489& 0.001& 6.7018& 0.001&46.601&0.01\\
2458754.90831&-40.511&0.829&0.4637&0.00105&0.25306&0.00112&0.34993&0.00147&0.34&8.99&7.11&1.93&0.0505& 0.001& 6.6993& 0.001& 46.564&0.01\\

 \\

\hline
\noalign{\vskip 0.5mm}

\hline

\end{tabular}}

\end{table}

\begin{table*}
\caption{Radial velocity and activity indices of HD\,33142 obtained from HIRES.} 
\label{tab:HIRES_RVdata}

\centering  

\begin{tabular}{@{}c c c c c c@{}} 

\hline\hline    
\noalign{\vskip 0.5mm}

Epoch [BJD] & RV [m\,s$^{-1}$] & $\sigma_{\rm RV}$ [m\,s$^{-1}$]  & $S$-index & H-index  \\  

\hline     
\noalign{\vskip 0.5mm}    

2454400.03215   &   -57.45   &    1.76 &   0.1150   &    0.03119        \\ 
2454461.87742   &   -23.62   &    1.72 &   0.1041   &    0.03108        \\ 
2454718.14941   &   -43.68   &    1.66 &   0.1167   &    0.03133        \\ 
2454806.95469   &   -13.93   &    3.11 &   0.1141   &    0.03114        \\ 
2454839.01455   &   11.14   &    1.68 &   0.1254   &    0.03118        \\ 
2454846.95912   &   6.44   &    1.96 &   0.1184   &    0.03110        \\ 
2454864.91463   &   33.49   &    1.56 &   0.1243   &    0.03113        \\ 
2454929.72406   &   6.19   &    1.65 &   0.1258   &    0.03117        \\ 
2455076.12029   &   -18.81   &    1.56 &   0.1208   &    0.03126        \\ 
2455085.08738   &   -16.46   &    1.49 &   0.1183   &    0.03131        \\ 
2455110.13017   &   -17.79   &    1.89 &   0.1172   &    0.03134        \\ 
2455173.05039   &   17.89   &    1.45 &   0.1244   &    0.03114        \\ 
2455187.90056   &   5.66   &    1.68 &   0.1284   &    0.03120        \\ 
2455188.95939   &   9.48   &    1.60 &   0.1188   &    0.03113        \\ 
2455189.82384   &   5.08   &    1.61 &   0.1197   &    0.03110        \\ 
2455190.89727   &   3.38   &    1.71 &   0.1220   &    0.03120        \\ 
2455196.80904   &   5.65   &    1.67 &   0.1186   &    0.03151        \\ 
2455197.96869   &   4.36   &    1.86 &   0.1222   &    0.03131        \\ 
2455198.99626   &   0.00   &    1.87 &   0.1280   &    0.03126        \\ 
2455255.74755   &   -2.69   &    1.31 &   0.1187   &    0.03132        \\ 
2455285.78066   &   -29.30   &    1.45 &   0.1332   &    0.03118        \\ 
2455312.72518   &   -43.10   &    1.79 &   0.0914   &    0.03110        \\ 
2455456.04311   &   -12.57   &    1.57 &   0.1238   &    0.03119        \\ 
2455490.95679   &   7.16   &    1.73 &   0.1237   &    0.03125        \\ 
2455521.96605   &   6.48   &    1.64 &   0.1176   &    0.03116        \\ 
2455546.06910   &   11.43   &    1.80 &   0.1125   &    0.03119        \\ 
2455584.91194   &   3.38   &    1.73 &   0.1202   &    0.03123        \\ 
2455633.81192   &   -28.14   &    1.45 &   0.1048   &    0.03120        \\ 
2455878.97668   &   17.34   &    1.68 &   0.1160   &    0.03128        \\ 
2455904.03568   &   7.60   &    1.51 &   0.1214   &    0.03112        \\ 
2455960.76706   &   -31.35   &    1.80 &   0.1188   &    0.03134        \\ 
2456018.74059   &   -57.46   &    1.61 &   0.1149   &    0.03127        \\ 
2456530.11307   &   13.05   &    1.58 &   0.1277   &    0.03129        \\ 
2456640.94484   &   -35.64   &    1.59 &   0.1137   &    0.03119        \\ 
2456913.14619   &   -39.74   &    1.49 &   0.1169   &    0.03120        \\ 
2457354.87263   &   -31.59   &    1.64 &   0.1153   &    0.03078        \\ 
2458715.13860   &   -43.22   &    1.50 &   -1.0000   &    -1.00000        \\ 
2458788.09457   &   1.08   &    1.55 &   -1.0000   &    -1.00000        \\ 
2458852.83152   &   15.02   &    1.54 &   -1.0000   &    -1.00000        \\ 
2458907.80135   &   -11.96   &    1.61 &   -1.0000   &    -1.00000        \\ 
  
\hline           
\end{tabular}

\end{table*}




\end{appendix}
\end{document}